\documentclass[10pt,journal,compsoc]{IEEEtran}

\usepackage{epsfig}
\usepackage{amssymb}
\usepackage{url}

\usepackage{subcaption}
\usepackage{amsmath}
\usepackage{amsthm}
\usepackage{amsfonts}
\usepackage{balance}

\usepackage[dvipsnames]{xcolor}
\usepackage{setspace}
\DeclareMathOperator*{\argmin}{arg\,min}

\newcommand{\rao}{r_{1\rightarrow 0}^{\mathbf{q},\mathbf{b}_i}}
\newcommand{\rat}{r_{0\rightarrow 1}^{\mathbf{q},\mathbf{b}_i}}

\usepackage{algorithmic}

\usepackage[utf8]{inputenc}
\usepackage{csvsimple}
\usepackage{tikz}
\usepackage{algorithm}
\usepackage{stmaryrd}
\usepackage{mathtools}
\usepackage{pgfplots}
\usepackage{filecontents}
\usepackage{tikz} 
\usetikzlibrary{shapes}
\usetikzlibrary{plotmarks}
\usepackage{comment}
\usepackage{multirow}
\usepackage{pgfplots}
\usepackage{amsmath,amssymb}

\usetikzlibrary{patterns}

\DeclarePairedDelimiter\ceil{\lceil}{\rceil}
\DeclarePairedDelimiter\floor{\lfloor}{\rfloor}

\newtheorem{example}{\textbf{Example}}
\theoremstyle{definition}
\newtheorem{definition}{\textbf{Definition}}

\usepackage{booktabs}
\usepackage{multirow}

\newenvironment{customlegend}[1][]{%
    \begingroup
    \csname pgfplots@init@cleared@structures\endcsname
    \pgfplotsset{#1}%
}{%
    \csname pgfplots@createlegend\endcsname
    \endgroup
}%

\def\addlegendimage{\csname pgfplots@addlegendimage\endcsname}

\pgfkeys{/pgfplots/number in legend/.style={%
        /pgfplots/legend image code/.code={%
            \node at (0.295,-0.0225){#1};
        },%
    },
}

\pgfplotsset{select coords between index/.style 2 args={
    x filter/.code={
        \ifnum\coordindex<#1\fi
        \ifnum\coordindex>#2\fi
    }
}}

\pgfplotsset{
    discard if not/.style 2 args={
        x filter/.code={
            \edef\tempa{\thisrow{#1}}
            \edef\tempb{#2}
            \ifx\tempa\tempb
            \else
                
            \fi
        }
    }
}

\makeatletter
\def\pgfmathfloatrounddisplaystyle@std#1#2e#3\relax{%
  \pgfmathfloatrounddisplaystyle@shared@impl#1#2e#3\relax{\times}{}{10^{#3}}%
}
\let\pgfmathfloatrounddisplaystyle\pgfmathfloatrounddisplaystyle@std
\makeatother

\begin{document}
\makeatletter
\title{Fast Cosine Similarity Search in Binary Space \\ with Angular Multi-index Hashing}

\author{Sepehr Eghbali and Ladan Tahvildari
\IEEEcompsocitemizethanks{\IEEEcompsocthanksitem S. Eghbali is with the Department of Electrical and Computer Engineering, University of Waterloo, 200 University Ave West, Waterloo, Ontario N2L 3G1, Email: s2eghbal@uwaterloo.ca.\protect\\

\IEEEcompsocthanksitem L. Tahvildari is with the Department of Electrical and Computer Engineering, University of Waterloo, 200 University Ave West, Waterloo, Ontario N2L 3G1, Canada. E-mail: ladan.tahvildari@uwaterloo.ca.}
\thanks{}
}

\IEEEtitleabstractindextext{%
\begin{abstract}
Given a large dataset of binary codes and a binary query point, we address how to efficiently find $K$ codes in the dataset that yield the largest cosine similarities to the query. The straightforward answer to this problem is to compare the query with all items in the dataset, but this is practical only for small datasets. One potential solution to enhance the search time and achieve sublinear cost is to use a hash table populated with binary codes of the dataset and then look up the nearby buckets to the query to retrieve the nearest neighbors. However, if codes are compared in terms of cosine similarity rather than the Hamming distance, then the main issue is that the order of buckets to probe is not evident. To examine this issue, we first elaborate on the connection between the Hamming distance and the cosine similarity. Doing this allows us to systematically find the probing sequence in the hash table. However, solving the nearest neighbor search with a single table is only practical for short binary codes. To address this issue, we propose the angular multi-index hashing search algorithm which relies on building multiple hash tables on binary code substrings. The proposed search algorithm solves the exact angular $K$ nearest neighbor problem in a time that is often orders of magnitude faster than the linear scan baseline and even approximation methods. 
\end{abstract}

}

\maketitle

\IEEEdisplaynontitleabstractindextext

\IEEEpeerreviewmaketitle

\IEEEraisesectionheading{\section{Introduction}\label{sec:introduction}}

\IEEEPARstart{R}{ecent} years have witnessed a surge of research on representing large-scale datasets with binary-valued features. Binary-valued representation has several advantages over real-valued representation, as binary vectors (codes) are more compact to store, faster to compare and cheaper to compute~\cite{muja2012fast}. 

Binary datasets are present in many research fields. In some applications, items of interest are described directly in terms of binary features. For example, in computer vision, many binary visual descriptors such as BRIEF~\cite{calonder2010brief}, BRISK~\cite{leutenegger2011brisk}, ORB~\cite{rublee2011orb}, and FREAK~\cite{alahi2012freak} have been proposed over the years. In the \emph{Bag of Words} model also, the absence or presence of words is used to quantify documents. However, perhaps the notable application of binary codes is in \emph{approximate nearest neighbor search} where high dimensional are encoded with \emph{compact} binary codes to enhance the search time~\cite{do2016learning,gong2013iterative,Liu_2016_CVPR, norouzi2011minimal,wang2014hashing,weiss2009spectral,zhang2013topology, Zhang_2016_CVPR}. This line of work has shown considerable promise for a variety of image and document search tasks such as near-duplicate detection~\cite{hajishirzi2010adaptive}, object detection~\cite{dean2013fast}, image retrieval~\cite{liu2012supervised} and pose estimation~\cite{shakhnarovich2003fast}.

One main advantage of incorporating binary codes is that the distance between two codes can be computed extremely fast using bitwise operators. For example, the Hamming distance between two codes can be computed by performing an XOR operation followed by counting the number of ones in the result (computed using the \emph{population count} operator). This important feature makes binary codes a suitable fit for the task of \emph{$K$ nearest neighbor search} ($K$NN) in which unseen test queries are compared against a large dataset to find the $K$ closest items. The $K$NN problem admits a straight forwards solution known as \emph{exhaustive search} or \emph{linear scan}: scanning the entire dataset and checking the distance of the query point to every code. Nevertheless, the cost of this approach is prohibitive for large scale datasets encountered in practice. Even if we use binary features to enhance the distance computation, the search time can still be in the order of minutes~\cite{ong2016improved}. Thus, it is imperative to obtain a solution with runtime that is \emph{sublinear} in the dataset size.

A relevant question concerns the possibility of utilizing data structures that provide sublinear search time, such as a hash table. It turns out that binary codes are in fact a suitable fit for hash tables as binary codes lie in a discrete space. To find $K$ nearest neighbors, a hash table is populated with binary codes where each code is treated as an index (memory addresses) in the hash table. Then, one can probe (check) the nearby buckets of the query point until $K$ items are retrieved. For instance, if the Hamming distance is used as the measure of similarity, then the algorithm that solves the exact $K$NN problem is as follows: starting with a Hamming radius equal to zero, $r=0$, at each step, the algorithm probes all buckets at the Hamming distance $r$ from the query. After each step, $r$ is increased by one, and the algorithm proceeds  until $K$ items are retrieved. However, in some applications, binary codes are compared in terms of cosine similarity, instead of the Hamming distance~\cite{bayardo2007scaling,Gong_angular:NIPS2012_4831,shrivastava2014defense}. This is known as the {\it angular $K$NN} problem. In such cases, there are no exact sequential procedures for finding the correct sequence of probings. In practice, instead of using a hash table, researchers resort either to the exhaustive search~\cite{Gong_angular:NIPS2012_4831}, or to the approximate similarity search techniques~\cite{shrivastava2014defense}, such as \emph{Locality Sensitive Hashing} (LSH)~\cite{Indyk:1998:ANN:276698.276876}.

In this paper, we propose a sequential algorithm for performing exact angular $K$NN search among binary codes. Our approach iteratively finds the sequence of hash table buckets to probe until $K$ neighbors are retrieved. We prove that, using the proposed procedure, the cosine similarity between the query and the sequence of generated buckets' indices will decrease monotonically. This means, the larger is the cosine similarity between a bucket index and the query, the sooner the index will appear in the sequence.

Using a hash table for searching can in principle reduce the retrieval time, nevertheless, this approach is only feasible for very compact codes, i.e., 32 bits at most~\cite{gong2013iterative, norouzi2012fast}. For longer codes (\emph{e.g.}, 64 bits), many of the buckets are empty and consequently the number of buckets that must be probed to find the $K$ nearest neighbors often exceeds the number of items in the dataset, making linear scan a faster alternative. {\it Multi-Index Hashing} (MIH)~\cite{aizawa1pqtable,greene1994multi,mao2016two, norouzi2014fast, norouzi2012fast, Ong_2016_CVPR} is a powerful technique for addressing this issue. The MIH technique hinges on dividing long codes into disjoint shorter codes to reduce the number of empty buckets. Motivated by the MIH technique proposed in~\cite{norouzi2012fast}, we develop the {\it Angular Multi-Index Hashing} (AMIH) technique to realize similar advantages for the angular $K$NN problem. Empirical evaluations of our approach show orders of magnitude improvement in search speed in conjunction with large-scale datasets in comparison with linear scan and approximation techniques.

Given a binary query and a hash table populated with binary codes, this study raises the following research questions which we address in the rest of the paper:
\begin{enumerate}
\item[RQ1:] What is correct probing sequence for solving the angular $K$NN problem?
\item[RQ2:] How can the MIH technique be tailored for the angular $K$NN problem?
\item[RQ3:] What is the effect of the AMIH technique on the query time?
\end{enumerate}

In a nutshell, we first establish a relationship between the cosine similarity and the Hamming distance. Relying on this connection, a fast algorithm for finding the correct order of probings is introduced. This allows modifying the multi-index hashing approach such that it can be applied to the angular $K$NN problem. 

\section{Related Works}
This section reviews some of the popular solutions to the nearest neighbor problem that relate to the current article. We first review earlier exact solutions to this problem. Then, we discuss more recent data-independent and data-dependent approximate solutions that can trade scalability versus accuracy. 

A classical paradigm to reduce the computational cost of the nearest neighbor search relies on tree-based indexing structures which offer logarithmic query time, $O(\log n)$, on the average. Perhaps the best known example of such techniques is the {\it kd-tree}~\cite{Bentley:1975:MBS:361002.361007} with a worst-case search time of $O(dn^{1-1/d})$. Following \cite{Bentley:1975:MBS:361002.361007}, many other tree-based indexing algorithms have been proposed (see~\cite{samet2006foundations} for an overview). However, these methods suffer from a phenomenon commonly known as the {\it curse of dimensionality}. This means, as the number of dimensions increases, these methods quickly reduce to the exhaustive search. Typically, the kd-tree and its variants are efficient for dimensions less than 20~\cite{Friedman:1977:AFB:355744.355745,weber1998quantitative}.

To overcome the apparent difficulty of devising algorithms that find the exact solution of the $K$NN problem, there has been an increasing interest to resort to {\it approximate} solutions.
Among the possible approximate techniques for solving the $K$NN problem, {\it Locality Sensitive Hashing} (LSH)~\cite{Indyk:1998:ANN:276698.276876} is perhaps the most notable which managed to break the linear query time bottleneck. The high-level idea of LSH is to use similarity preserving hash functions such that, with a high probability, points that are near to each other are mapped to the same hash bucket.

To find {\it compact} codes, instead of using random projections, recent studies have aimed at finding data dependent mappings to reduce the number of required bits.  Salakhutdinov and Hinton~\cite{salakhutdinov2009semantic} used {\it Restricted Boltzmann Machine} as an auto-encoder to learn the underlying binary codes. Weiss et al.~\cite{weiss2009spectral} have proposed an eigenvector formulation which aims at finding short similarity preserving binary codes, such that bits are uncorrelated and balanced. {\it Binary Reconstructive Embedding} (BRE)~\cite{kulis2009learning} uses a loss function that penalizes the difference between the Euclidean distance in the input space and the Hamming distance in the binary space. More recent techniques rely on using neural networks and non-linear hash functions to better preserve some notion of similarity~\cite{carr2015, Liu_2016_CVPR, Zhang_2016_CVPR}.

While Hamming distance is the most popular measure of similarity used to compare binary codes~\cite{EghbaliAT17,norouzi2012fast}, in some applications, codes are compared in terms of cosine similarities. For example, Gong et al.~\cite{Gong_angular:NIPS2012_4831} have developed a binary hashing technique in which the resulting codes are compared in terms of their corresponding angles. Also, in Bag of Words representation, it is common to compare the codes with respect to the cosine similarity measure~\cite{shrivastava2014defense}. Researchers have also proposed other distance measures for binary codes such as spherical Hamming distance~\cite{heo2012spherical} and Manhattan distance~\cite{kong2012manhattan}.

The idea of using hash tables to avoid exhaustive search in \emph{Approximate Nearest Neighbor} (ANN) techniques has been studied in recent years. Liu et al.~\cite{liu2013reciprocal} proposed an algorithm for partitioning a pool of hash functions into sets such that the hash functions of each set are independent of each other. Then, they use each set of hash functions to form a separate hash table. Babenko et al.~\cite{babenko2012inverted} have proposed the {\it inverted multi-index} technique to solve the ANN problem for problems in which the distance between data points are estimated with codewords of multiple codebooks. This technique creates two hash tables by decomposing data vectors into two disjoint substrings and hashing each substring in one of the tables. The query is similarly decomposed into two substrings and the search is performed in each hash table to find the corresponding nearest neighbor. More recently, Matsui et al.~\cite{aizawa1pqtable} have used multiple hash tables to reduce the search time, whereby the distance between items are approximated by codewords of multiple codebooks. Iwamura et al.~\cite{iwamura2013most} have also proposed a non-exhaustive search algorithm based on the branch and bound technique.

\section{Definitions and Problem Statement}
\label{sec::background}
Table~\ref{tab::notation} shows the notations used throughout this paper. Some further notations will be defined in Section~\ref{sec::fast}.

Given the dataset $\mathcal{B} = \{\mathbf{b}_i\in \{0,1\}^p\}_{i=1}^n$, and the query $\mathbf{q}\in \{0,1\}^p$, the aim of the (binary) Nearest Neighbor (NN) problem, also called the 1 Nearest Neighbor (1NN)  problem, is to find the item in $\mathcal{B}$ that is the closest to $\mathbf{q}$:
\begin{equation}
\text{NN}(\mathbf{q}) = \argmin_{\mathbf{b}\in \mathcal{B}} dis(\mathbf{q},\mathbf{b}),
\end{equation}
where $dis(.,.)$ is the distance between two items. The $K$ Nearest Neighbor problem ($K$NN) is the generalization of 1NN, aiming to find the $K$ closest items to the query.
When $dis$ is the Euclidean distance, the problem is called the Euclidean $K$NN, and when $dis$ is the inverse of the cosine similarity, it is called the angular $K$NN. The focus of this paper is to efficiently solve the angular $K$NN in conjunction with datasets of binary codes.

Another related search problem is the $R$-near neighbor problem ($R$NN). The goal of $R$NN problem to report all data points lying at distance at most $R$ from the query point. Similarly, if the Euclidean distance is used as the similarity measure, we call the problem the Euclidean $R$-near neighbor.

$R$NN and $K$NN problems are closely related. For binary datasets, one way to tackle the Hamming $K$NN problem is to solve multiple instances of the Hamming $R$NN problem. First, a hash table is populated with binary codes in $\mathcal{B}$. Then, starting from a Hamming radius equal to zero, $R=0$, the procedure increases $R$ and then solves the $R$-near problem by searching among the buckets at the Hamming distance $R$ from the query. This procedure iterates until $K$ items are retrieved. Nevertheless, if cosine similarity is used, the probing sequence will not be the same as the case of the Hamming distance. Unlike the Hamming distance, the angle between two binary codes is not a monotonically increasing function of their Euclidean distance. In other words, if binary codes $\mathbf{b}_1$ and $\mathbf{b}_2$ satisfy $\lVert\mathbf{q}-\mathbf{b}_1\rVert_H >\lVert\mathbf{q}-\mathbf{b}_2\rVert_H$, it does not necessarily lead to $sim(\mathbf{q},\mathbf{b}_1)<sim(\mathbf{q},\mathbf{b}_2)$ where $sim(\mathbf{x},\mathbf{y})$ is the cosine of the angle between binary codes $\mathbf{x}$ and $\mathbf{y}$, and $\lVert.\rVert_H$ denotes the Hamming norm. Next, we propose an algorithm that systematically finds the order of probings required for solving the angular $K$NN problem.

\begin{table}[]
\centering
\caption{Notations}
\label{tab::notation}
\begin{tabular}{|c|c|}\hline
\textbf{Symbol}              & \textbf{Explanation} \\ \hline
 $K$                         & Number of nearest neighbors to retrieve \\ \hline
 $\mathbf{q}$                &     Binary query vector \\ \hline
 $\lVert\mathbf{.}\rVert_u$            & $\ell_u$ norm     \\ \hline
 $p$                         & Length of the binary codes \\ \hline
 $\mathcal{H}_{\mathbf{q},\mathbf{b}_i}$ & Hamming distance tuple between $\mathbf{q}$ and $\mathbf{b}_i$ \\ \hline
 $sim(\mathbf{q},\mathbf{b}_i)$    & Cosine of the angle between $\mathbf{q}$ and $\mathbf{b}_i$\\ \hline
 $\rat$                           & \# of bits that are zero in $\mathbf{q}$ and one in $\mathbf{b}_i$ \\ \hline
 $\rao $							& \# of bits that are one in $\mathbf{q}$ and zero in $\mathbf{b}_i$ \\ \hline
 $\lVert.\rVert_H$   & Hamming norm \\ \hline
 $\mathcal{B}$                    & Dataset of binary codes \\ \hline
\end{tabular}
\end{table}

\section{Fast Cosine Similarity Search}
\label{sec::fast}

To reduce the search cost, we propose to use a hash table populated with binary codes. Given the dataset $\mathcal{B}$, we populate a hash table with items of $\mathcal{B}$, where each binary code is treated as the direct index of a hash bucket. The problem that we aim to tackle is finding the $K$ closest binary codes (in terms of cosine similarity) to the query. Evidently, for a given query $\mathbf{q}$, the binary code that yields the largest cosine similarity is $\mathbf{q}$ itself. Therefore, the first bucket to probe has the index identical to $\mathbf{q}$. The next bucket to probe has the second largest cosine similarity to $\mathbf{q}$, and so on. In the rest of this section, we propose an algorithm for efficiently finding such a sequence of probings to address our first research question (RQ1).

The cosine similarity of two binary codes can be computed using:

\begin{equation}
\label{eq::cosine_sim}
sim(\mathbf{q},\mathbf{b}_i) = \frac{\langle\mathbf{q}, \mathbf{b}_i\rangle}{\lVert\mathbf{q}\rVert_2\lVert\mathbf{b}_i\rVert_2},
\end{equation}
where $\langle \cdot, \cdot\rangle$ denotes the inner product and $\lVert\cdot\rVert_u$ denotes the $\ell_u$ norm.
In comparison to the Hamming distance, computing the cosine similarity is computationally more demanding. While computing Hamming distance requires an XOR followed by popcount operator, calculating cosine similarity needs the square roots and a division, refer to~(\ref{eq::cosine_sim}).

The key idea behind our technique relies on the fact the set of all binary codes at the Hamming distance $r$ from the query can be partitioned into $r+1$ subsets, where the codes in each subset yield equal cosine similarities to the query. In particular, for two binary code $\mathbf{q}$ and $\mathbf{b}_i$ lying at Hamming distance $r$ from each other, there are $r$ bits that differ in the two vectors. Let $\rao$ denote the number of bit positions that are 1 in $\mathbf{q}$ and 0 in $\mathbf{b}_i$. Similarly, let $\rat$ denote the number of bit positions that are 0 in $\mathbf{q}$ and 1 in $\mathbf{b}_i$.  By the definition of Hamming distance, we have $\rao+\rat=r$. Consequently, we can rewrite (\ref{eq::cosine_sim}):
\begin{equation}
\label{eq:cosine_r1r2}
sim(\mathbf{q},\mathbf{b}_i) = \frac{\lVert\mathbf{q}\rVert_1-\rao}{\sqrt{\lVert\mathbf{q}\rVert_1} \times \sqrt{\lVert\mathbf{q}\rVert_1-\rao+\rat}}.
\end{equation}
It is clear that $0\leq\rao\leq||\mathbf{q}||_1$ and $0\leq\rat\leq p-||\mathbf{q}||_1$. 

The dot product of two binary codes (the numerator of~(\ref{eq::cosine_sim})) is equal to the number of positions where $\mathbf{q}$ and $\mathbf{b}_i$ are both 1, which is equal to $\lVert\mathbf{q}\rVert_1-\rao$. The denominator simply contains the $\ell_2$ norms of $\mathbf{q}$ and $\mathbf{b}_i$. In the rest of this paper, we use~(\ref{eq:cosine_r1r2}) to compute the cosine similarity. 

The important observation is that, for a given query $\mathbf{q}$, all binary codes which correspond to the same values of $r_1$ and $r_2$ lie at the same angle from $\mathbf{q}$. We use this observation to define the notation of~\emph{Hamming Distance Tuple} as follows:

\begin{definition} \textsc{(Hamming Distance Tuple)} \normalfont Given a query $\mathbf{q}$, we say a given binary code $\mathbf{b}_i$ lies at the Hamming distance tuple $\mathcal{H}_{\mathbf{q},\mathbf{b}_i}=(\rao,\rat)$ from $\mathbf{q}$ if:
\begin{enumerate}
\item[a)] the number of bit positions in which $\mathbf{q}$ is 1 and $\mathbf{b}_i$ is 0 equals $\rao$, and,
\item[b)] the number of bit positions in which $\mathbf{q}$ is 0 and $\mathbf{b}_i$ is 1 equals $\rat$.
\end{enumerate}
\end{definition}
A Hamming distance tuple, such as $(r_1,r_2)$, is \emph{valid} if both of its elements are in valid ranges, i.e., $0\leq r_1 \leq \lVert\mathbf{q}\rVert_1$ and $0 \leq r_2 \leq p-\lVert\mathbf{q}\rVert_1$.

Each Hamming distance tuple represents a set of binary codes lying at the same angle from $\mathbf{q}$. The number of binary codes lying at the Hamming distance tuple $(\rao,\rat)$ from $\mathbf{q}$ is:

\begin{equation}
\label{eq:numberofbuckets}
 \binom{\lVert\mathbf{q}\rVert_1}{\rao} \times \binom{p-\lVert\mathbf{q}\rVert_1}{\rat}.
\end{equation}
As all codes with the same Hamming distance tuple yield identical $sim$ values, instead of searching for the correct probing sequence, we find the correct sequence of Hamming distance tuples.

We say that a Hamming distance tuple $(r'_1,r'_2)$ is less than or equal to $(r_1,r_2)$, shown by $(r'_1,r'_2)\preceq(r_1,r_2)$, if and only if $r'_1\leq r_1$ and $r'_2\leq r_2$.

\begin{definition} \textsc{($(r_1,r_2)$)-near neighbor) }\normalfont A binary code $\textbf{b}_i$ is called an $(r_1,r_2)$-near neighbor of $\textbf{q}$, if we have $\mathcal{H}_{\mathbf{q},\mathbf{b}_i} \preceq (r_1,r_2)$.
\end{definition}

\begin{example} \normalfont Suppose $\mathbf{q}=(1, 1, 1, 0, 0, 0)$ and $\mathbf{b}_1=(0,1,0,1,1,1)$, then $\mathbf{b}_1$ lies at the Hamming distance tuple $\mathcal{H}_{\mathbf{q},\mathbf{b}_1}=(2,3)$ from $\textbf{q}$, and $\mathbf{b}_2 = (1, 1, 1, 1, 1, 1)$ lies at the Hamming distance tuple $\mathcal{H}_{\mathbf{q},\mathbf{b}_2}=(0,3)$ from $\mathbf{q}$. Also, the Hamming distance tuple $(0,3)$ is less than the Hamming distance tuple $(2,3)$.
\end{example}

The partial derivatives of~(\ref{eq:cosine_r1r2}) with respect to $\rao$ and $\rat$ are both negative. This property indicates that, for a given Hamming distance tuple $(x,y)$, all the binary codes with the Hamming distance tuple $(x',y')$ satisfying $(x',y')\preceq (x,y)$ have larger $sim$ values.

\begin{figure}
\centering
\begin{tikzpicture}[scale=0.7]
\begin{axis}[
    xlabel=$\rao$, ylabel = $\rat$, zlabel = cos,
    view={140}{80},
        ztick={0,1},
    colorbar
]
\addplot3[only marks,
    scatter,
    samples = 32,
    samples y=14,
    domain = 0:31,
    domain y = 0:13,
]
{(32-x)/(sqrt(32)*sqrt(32-x+y))};
\end{axis}
\end{tikzpicture}
\caption{Plot of $sim$ values for different values of $\rao$ and $\rat$ with $p=45$ and $\lVert\mathbf{q}\rVert_1=32$.}
\label{fig::plot}
\end{figure}
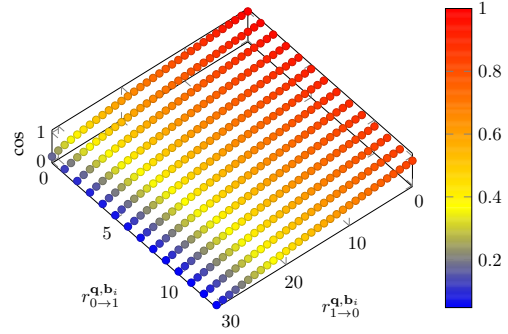

To visualize how the value of $sim$ varies with respect to $\rao$ and $\rat$, the $sim$ value as a function of $\rao$ and $\rat$ is plotted in Fig.~\ref{fig::plot}. We are interested in sorting the tuples $(\rao,\rat)$ (small circles in Fig.~\ref{fig::plot}) in decreasing order of $sim$ values. A naive way to construct the probing sequence is to compute and sort the $sim$ values of all possible tuples. However, in a real application, we expect to use a small fraction of the Hamming distance tuples as we only need to probe the hash buckets until $K$ neighbors are retrieved. Next, we propose an efficient algorithm that, in most cases, requires neither sorting, nor computing the $sim$ values. 

\begin{definition} \textsc{(Hamming Ball)} \normalfont For a given query $\mathbf{q}$, the set of all binary codes with a Hamming distance of at most $r$ from $\mathbf{q}$ is called the Hamming ball centered at $\mathbf{q}$ with radius $r$, and is shown by $\mathcal{C}(\mathbf{q},r)$:
\begin{equation}
\mathcal{C}(\mathbf{q},r)=\{\mathbf{h}\in\{0,1\}^p: \lVert\mathbf{q}-\mathbf{h}\rVert_H\leq r\}.
\end{equation}
\end{definition}

Given $\mathbf{b}_i$, values of $\rao$ and $\rat$ can be computed efficiently using bitwise operations. However, to search for the $K$ closest neighbors in the populated hash table, we are interested in progressively finding the values of $\rao$ and $\rat$ that lead to binary codes with the largest $sim$ value. One observation is that, within all indices lying at the Hamming distance $r$ from $\mathbf{q}$, indices with the Hamming distance tuples $(\rao,\rat) = (0,r)$ and $(\rao,\rat) = (r,0)$ (provided that they are valid tuples) yield the largest and the smallest cosine similarities with the query, respectively. This fact leads to the following proposition:

{\it \textbf{Proposition 1}: Among all binary codes lying at the Hamming distance $r$ from $\mathbf{q}$, those with larger values of $\rat$ yield larger $sim$ values.}

{\it Proof:} To prove this proposition, let us compute the derivatives of~(\ref{eq:cosine_r1r2}) with respect to $\rat$ or $\rao$ (in this proposition, to be able to take derivatives, we assume that $\rat$ and $\rao$ are continuous variables in $\mathbb{R}^+$). Suppose $\rao + \rat = r$, by replacing $\rao$ with $r-\rat$, we obtain:

\begin{equation}
\label{eq:fixedR}
sim(\mathbf{q},\mathbf{b}_1) = \frac{\lVert\mathbf{q}\rVert_1+\rat-r}{\sqrt{\lVert\mathbf{q}\rVert_1}\times\sqrt{\lVert\mathbf{q}\rVert_1+2\rat-r}}.
\end{equation}
After some algebraic manipulations, it follows that $\frac{\partial sim}{\partial \rat}\geq0$. Therefore, among all tuples at the Hamming distance $r$ from $\mathbf{q}$, the maximum of $sim$ occurs at $(\rao,\rat) = (0,r)$, and its minimum occurs at $(\rao,\rat)=(r,0)$.$\blacksquare$

Proposition 1 states that, for tuples at the Hamming distance $r$ from $\mathbf{q}$, $sim$ is a growing function of $\rat$. As a result, the order of tuples in the direction of decreasing $sim$ values is $(0,r),(1,r-1),\ldots,(r,0)$. In other words, among all the binary codes that lie at the Hamming distance $r$ from the query, those with larger $\ell_1$ norms yield larger cosine similarities. 

While the Proposition 1 specifies the direction of the search for a given Hamming distance, it does not establish the relationship between the Hamming distance and the cosine similarity for different Hamming distances.

Although the above may appear as a discouraging observation, we show that, for small Hamming distances, the cosine similarity and the Hamming distance are related to each other. In particular, the following proposition specifies the region where the cosine similarity is a monotonically decreasing function of the Hamming distance.

{\it \textbf{Proposition 2}: If $\lVert\mathbf{q}\rVert_1>\frac{r(r+t)}{t}$ for some $r,t\in\{1,\ldots,p\}$, then all binary codes in $\mathcal{C}(\mathbf{q},r)$ yield larger cosine similarities to $\mathbf{q}$ than binary codes with Hamming distances at least $r+t$ from $\mathbf{q}$.}

{\it Proof:} According to Proposition 1, the maximum of the $sim$ value for a fixed Hamming distance $r$ occurs at $(\rao,\rat)=(0,r)$ with a $sim$ value of $\sqrt{\frac{z}{z+r}}$, and its minimum occurs at $(\rao,\rat)=(r,0)$ with a $sim$ value of $\sqrt{\frac{z-r}{z}}$, where $z=\lVert\mathbf{q}\rVert_1$. The condition in Proposition 2 is satisfied if the smallest value of $sim(\mathbf{q},\mathbf{b}_i)$, where $\mathbf{b}_i \in \mathcal{C}(\mathbf{q},r)$, is larger than the largest value of $sim(\mathbf{q},\mathbf{b}_j)$ where $\mathbf{b}_j$ lies at the Hamming distance $r+t$ from $\mathbf{q}$. Hence, we have:
\begin{equation}
\begin{split}
& \sqrt{\frac{\lVert\mathbf{q}\rVert_1-r}{\lVert\mathbf{q}\rVert_1}} > \sqrt{\frac{\lVert\mathbf{q}\rVert_1}{\lVert\mathbf{q}|\rVert_1+r+t}} \\ \\ 
& \Rightarrow \frac{\lVert\mathbf{q}\rVert_1-r}{\lVert\mathbf{q}\rVert_1} > \frac{\lVert\mathbf{q}\rVert_1}{\lVert\mathbf{q}\rVert_1+r+t} \\ \\
& \Rightarrow (\lVert\mathbf{q}\rVert_1-r)(\lVert\mathbf{q}\rVert|_1+r+t)> \lVert\mathbf{q}\rVert_1^2 \\ \\
& \Rightarrow \lVert\mathbf{q}\rVert_1>\frac{r(r+t)}{t}.
\end{split}
\end{equation}
This concludes the proof.$\blacksquare$

If the condition of Proposition 2 is satisfied for $t=1$ and some radius of search $r$, then all the binary codes inside the Hamming ball $\mathcal{C}(\mathbf{q},r)$ have larger cosine similarities than those outside of $\mathcal{C}(\mathbf{q},r)$. Also, among all binary codes inside $\mathcal{C}(\mathbf{q},r)$, those with larger Hamming distances from the query have smaller cosine similarities.That is, if $\mathbf{b}_i$ is closer to $\mathbf{q}$ than to $\mathbf{b}_j$ in terms of the Hamming distance, then $\mathbf{b}_i$ is also closer to $\mathbf{q}$ in terms of the cosine similarity.

Therefore, for binary codes lying within the Hamming ball $\mathcal{C}(\mathbf{q},r)$, cosine similarity is a decreasing function of the Hamming distance. In this case, the search algorithm is straightforward: for $t=1$, the maximum integer $r$ that satisfies the inequality condition in Proposition 2 is found. Let $\hat{r}$ denote the integer part of the positive root of the equation $r^2+r-\lVert \mathbf{q}\rVert_1$ (this equation has only one positive root). Staring from $r=0$, the search algorithm increases the Hamming radius until the specified number of neighbors are retrieved, or until $r$ reaches $\hat{r}$. Further we know that, for each Hamming radius, the search direction should be aligned with the direction specified by the Proposition 1. 

\begin{definition} \normalfont ($(r_1,r_2)$\textsc{-near Neighbor Problem}) Given the query point $\mathbf{q}$ and dataset $\mathcal{B}$, the result of $(r_1,r_2)$-near neighbor problem is the set of all codes in $\mathcal{B}$ that lie at a Hamming distance tuple of at most $(r_1,r_2)$ from $\mathbf{q}$. 
\end{definition}

Our approach, is effective in cases that $K$ binary codes are retrieved before $r$ reaches $\hat{r}$. It tackles the angular $K$NN problem by solving multiple instances of the $(r_1,r_2)$-near neighbor problem. An important advantage of the proposed algorithm is that it does not need to compute the actual $sim$ values between binary codes. It can be efficiently implemented using bitwise operators and the popcount function.  In the rest of this section, we address the case of $r>\hat{r}$.

When the search radius is greater than $\hat{r}$, the $sim$ value is not a monotonically decreasing function of the Hamming distance. However, we propose a sequential algorithm that can efficiently find the proper ordering of the tuples. Our key idea is that, although the next tuple in the ordering can lie at many different Hamming distances, we show that it can be found by searching among a small subset of remaining tuples. In particular, we first form a small set of candidate tuples and then select the one with the highest $sim$ value. To do that, one can first insert such candidate tuples (which we call them \emph{anchors}) into a priority queue and then sequentially select the one with highest priority. The priority of a tuple is evidently determined by its corresponding $sim$ value. The queue is initialized with the tuple $(0,\hat{r}+1)$. When a tuple is pushed into the queue, it is considered as traversed. At each subsequent step, the tuple with the top priority (the highest $sim$ value) is popped from the queue. When a tuple is popped, two tuples are considered for insertion into the  queue.  Hereafter, these are called  {\it the first anchor} and {\em the second anchor}, respectively. These two tuples are checked, and if ``valid" and ``not traversed", they are pushed into the queue. 

\begin{definition} \textsc({First and Second Anchors of a Tuple)}

\normalfont Given a query $\mathbf{q}$ and a Hamming distance tuple $R=(x,y)$, the first anchor and the second anchor of $R$ are defined as follows:
\begin{itemize}

 \item Among all tuples that lie at the Hamming distance $x+y+1$ from $\mathbf{q}$, the tuple with the largest $sim$ value is called the {\it first anchor} of $R$. 
 
 \item Among all tuples that lie at the Hamming distance $x+y$ from $\mathbf{q}$ and have smaller $sim$ values than $R$, the tuple with the largest $sim$ value is called the {\it second anchor} of $R$.
 \end{itemize}
\end{definition}

\begin{example} \normalfont For the query $\mathbf{q}$ with $\lVert\textbf{q}\rVert_1=10$ and $p=32$, the first anchor of  $v=(1,4)$ is $(0,6)$ and the second anchor is $(2,3)$ (according to the Proposition 1).
\end{example}

When a tuple is popped from the queue, the algorithm pushes the first and the second anchors of the popped tuple into the priority queue (provided that these are valid, and not traversed) and marks them as traversed. This procedure continues until either $K$ elements are retrieved, or all valid tuples are traversed. Therefore, when a tuple such as $R=(x,y)$ is popped from the queue, the algorithm checks whether the following two tuples are valid or not:

\begin{enumerate}
\item[a)] {\bf The first anchor of $R$}: This tuple, by definition, has the largest $sim$ value among the tuples at the Hamming distance $x+y+1$ from the query. According to Proposition 1, this candidate is $(0,x+y+1)$ if $x+y+1\leq p-\lVert\mathbf{q}\rVert_1$. In general, to ensure that the two components of this tuple are in acceptable ranges, the first anchor of $R$ takes the form $(c,x+y+1-c)$ where $c=\max(0,x+y+1-(p-\lVert\mathbf{q}\rVert_1))$. Note that the first component of any Hamming distance tuple is at most $\lVert\mathbf{q}\rVert_1$ (number of ones in $\mathbf{q}$) and its second component is at most $p-\lVert\mathbf{q}\rVert_1$ (number of zeros in $\mathbf{q}$).
\item[b)] {\bf The second anchor of $R$}: Among the tuples that have smaller $sim$ values than $R$, and lie at the Hamming distance $x+y$ from the query, this tuple is the one that has the largest $sim$ value. Using Proposition 1, it is easy to show that the second anchor of $R$ is $(x+1,y-1)$. This tuple is pushed into the queue if its components are in acceptable ranges (the second anchor is valid if $x+1\leq\lVert\mathbf{q}\rVert_1$ and $y-1\geq 0$).

\end{enumerate} 

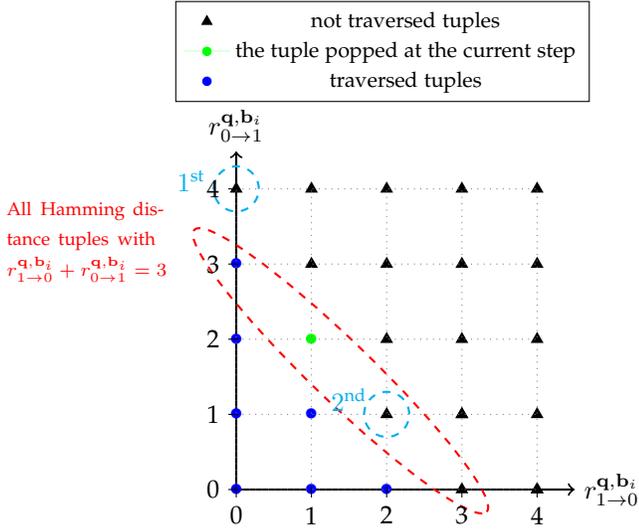
\begin{figure}
\begin{tikzpicture}[x=1cm,y=1cm]
  \draw [<->,thick] (0,4.5) node (yaxis) [above] {$\rat$}
        |- (4.5,0) node (xaxis) [right] {$\rao$};

\draw	(-1.5,3.3) node[text width=3.1cm,red]{{\scriptsize All Hamming distance tuples with $\rao+\rat=3$}};

\foreach \Point in {(0,0), (0,1), (1,0), (2,0), (1,1),(0,2),(0,3)}{
    \node [blue] at \Point {\textbullet};
    }
\foreach \Point in {(3,1), (4,0),  (1,4), (2,3), (3,2), (4,1), (2,4), (3,3), (4,2), (3,1),(4,0),(0,4),(1,3),(2,1),(2,2),(3,0),(4,4),(3,4),(4,3)}{
    \node [fill=black, regular polygon, regular polygon sides=3,inner sep=1pt] at \Point {};
    }

\node [green] at (1,2) {\textbullet};

\draw[red,thick,dashed,rotate=-44] (-0.1,2.1) ellipse (2.7cm and 0.4cm);
\draw[cyan,thick,dashed] (0,4) circle (0.3cm);
\draw[cyan,thick,dashed] (2,1) circle (0.3cm);

\draw [dotted, gray] (0,0) grid (4,4);

 	\foreach \x in {0,...,4}
     		\draw (\x,1pt) -- (\x,-3pt)
			node[anchor=north] {\x};
    	\foreach \y in {0,1,...,4}
     		\draw (1pt,\y) -- (-3pt,\y) 
     			node[anchor=east] {\y}; 

\node [cyan] at (1.5,1.2){$2^\text{nd}$};
\node [cyan] at (-0.6,4.1){$1^\text{st}$};
\begin{customlegend}[
legend entries={
not traversed tuples,
the tuple popped at the current step,
traversed tuples
},
legend style={at={(4.7,6.5)},font=\footnotesize}]
    \addlegendimage{mark=triangle*, mark size=3pt,draw=white}
    \addlegendimage{mark=*,color=green, fill=green,draw=white}
    \addlegendimage{mark=*,color=blue,draw=white}
\end{customlegend}
\end{tikzpicture}
\caption{Visual representation of the ``first anchor" and the ``second anchor" of a Hamming distance tuple.}
\label{fig::anchor}
\end{figure}

Fig.~\ref{fig::anchor} shows an example of the first/second anchors (shown in dashed circles) of a tuple that is selected in the current step (shown in green).

Next, we prove that the proposed algorithm results in the correct ordering of Hamming distance tuples. 

{\it \textbf{Proposition 3}: In each iteration, the Hamming distance tuple popped from the queue has a smaller $sim$ value than the traversed tuples, and has the largest $sim$ value among the not traversed tuples. Moreover, the algorithm eventually traverses every tuple.}

{\it Proof}: When $r<\hat{r}$, according to Propositions 1 and 2, the ordering is correct. For $r\geq \hat{r}$, we show that the selected candidate has the highest cosine similarity among the remaining tuples.

Assume that the algorithm is not correct. Let $R$ be the first tuple that the algorithm selects incorrectly. This means another tuple, such as $R'=(x',y')$, yields the highest $sim$ value and it has not been pushed into the priority queue because if $R'$ had been pushed into the queue, then  $R'$ would have been popped from the queue instead of $R$. Let $r'=x'+y'$, meaning that $r'$ is the Hamming distance between $\mathbf{q}$ and any binary code lying at the Hamming distance tuple $(x',y')$ from the $\mathbf{q}$. Consider all binary codes that lie at the Hamming distance $r'$ from $\mathbf{q}$. If there exists a tuple with the second component greater than $y'$ that has not been traversed yet, then a contradiction occurs (this means $R'$ does not yield the largest $sim$ value). This stems from the fact that, at a fixed Hamming distance, tuples with larger second components have larger $sim$ values (Proposition 1). As a result, $y'$ yields the largest possible value among the not-traversed tuples lying at the Hamming distance $r'$ from $\mathbf{q}$. However, we show that, this tuple should have been pushed into the priority queue in previous steps. One of the following cases may occur:

\begin{enumerate}
\item[a)] Until the current step, no Hamming distance tuple at the Hamming distance $r'$ from $\mathbf{q}$ has been selected: According to Proposition 1, any tuple with the Hamming distance $r'-1$ that is in the set $\mathcal{L}=\{(a,b)|(a,b) \text{ is a valid tuple and, } a+b=r'-1,a\leq x', b\leq y \}$ has larger $sim$ values than $R'$. Therefore, all of them must have been selected prior to $R'$ in the sequence. However, the first time that a tuple from $\mathcal{L}$ was popped, $R'$ was pushed into the priority queue. $R'$ is in fact the first anchor of all the tuples in $\mathcal{L}$, and thus, it must have been pushed when any of the elements in $\mathcal{L}$ were popped from the priority queue.

\item[b)]  At least one Hamming distance tuple with the Hamming distance $r'$ from $\mathbf{q}$ has been traversed in previous steps: Similar to the previous case, $R'$ was pushed into the priority queue when the algorithm popped the tuple $(x'-1,y'+1)$. In this scenario, $R'$ is the second anchor of $(x'-1,y'+1)$.
\end{enumerate}
It is concluded that $R'$ must have been pushed into the priority queue during previous steps, which contradicts the assumption that $R'$ is not a member of the priority queue.

We also need to prove that the algorithm is {\it complete}, i.e., the algorithm continues until it either finds the $K$ neighbors, or it traverses all the valid tuples. Again, let us assume the contrary. This means that, at the final step, the algorithm pops the last tuple from the queue and the last tuple does not have any valid anchors. Thus, the queue remains empty and the algorithm will terminate while there are still some valid tuples that have not been traversed. It is clear that the not-traversed tuples cannot lie at the Hamming distance of $r$ when at least one tuple with the Hamming distance $r$ is traversed. This situation occurs because once the first tuple with the Hamming distance $r$ is popped from the queue, the second anchor of this tuple is pushed into the queue. Therefore, one tuple with the Hamming distance $r$ always exists in the queue until the last one of such tuples is popped, and such a last tuple does not have a valid second anchor. As a result, the only possible case is that all the tuples at a Hamming distance less than or equal to $r$ have been traversed; and all of the tuples at a Hamming distance $r+1$ and greater have not been traversed. However, this is not possible because when a tuple at the Hamming distance $r$ is popped from the queue, its first anchor is pushed into the queue and this tuple lies the Hamming distance $r+1$. Hence, all the tuples at the Hamming distance $r+1$ from the query will be covered eventually.$\blacksquare$

\section{Angular Multi-index Hashing}

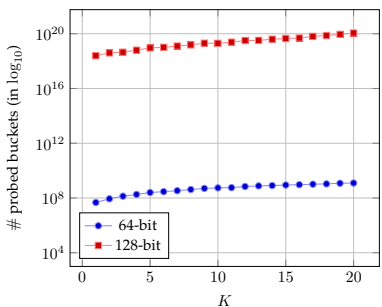
\begin{figure}[b]
\centering
\begin{tikzpicture}[scale=0.6]
\begin{axis}[
grid = both,
xlabel={$K$},
ylabel={$\#$ probed buckets (in $\log_{10}$)},
ymin=1000,
ymode=log,
legend pos=south west
]
\addplot+[color=blue!50]table [x=kn,y=num,col sep=comma]{bucketschecked64.csv};
\addlegendentry{64-bit}
\addplot+[color=red!50]table [x=kn,y=num,col sep=comma]{bucketschecked128.csv};
\addlegendentry{128-bit}
\end{axis}
\end{tikzpicture}
\caption{The average number of probings required for solving the angular $K$NN problem, if a single hash table is used for the SIFT dataset (with $10^9$ items).}
\label{fig::bucketstocheck}
\end{figure}

To achieve satisfactory retrieval accuracy, applications of binary hashing often require binary codes with large lengths (e.g., 64 bits).  For such applications, it is not practical to use a single hash table mainly because of the computational cost of search. For long binary codes, it is frequently the case that $n \ll 2^p$ and thus most of the buckets in the populated hash table are empty. To solve the $K$NN problem in such a sparse hash tables, even for small values of $K$, often the number of buckets to be examined exceeds the number of items in the dataset. This means that the exhaustive search (linear scan) is a faster alternative than using a hash table. As shown in Fig.~\ref{fig::bucketstocheck}, the average number of probing required for solving the angular $K$NN query for the SIFT dataset with one billion items (the details of SIFT will be explained later), often exceeds the number of available binary codes in the dataset.
This problem arises as the required number of probings grows near-exponentially with the values of $\rao$ and $\rat$ (refer to~(\ref{eq:numberofbuckets})).

Multi-Index Hashing (MIH)~\cite{greene1994multi}, and its variants~\cite{norouzi2014fast, norouzi2012fast}, are elegant approaches for reducing storage and computational costs of the $R$-near neighbor search for binary codes. The key idea behind the multi-index hashing is that, as many of the buckets are empty, one can merge the buckets over different dimensions of the Hamming space. To do this, instead of creating one huge hash table, MIH creates multiple smaller hash tables with larger buckets, where each bucket may be populated with more than one item. To do this, all binary codes are divided into smaller disjoint (usually with the same length) substrings, then each substring is indexed within its corresponding hash table. Therefore instead of one creating one huge hash table, the idea of MIH is to form multiple smaller hash tables which can significantly reduce the storage cost.

More importantly, MIH reduces the computational cost of the search. To solve the $R$-near neighbor problem, the query is similarly partitioned into $m$ substrings. Then, MIH solves $m$ instances of the $\frac{R}{m}$-near neighbor problem, one per each hash table. By doing this, the neighbors of each substring in its corresponding hash table are retrieved to form a set of potential neighbors. Since some of the retrieved neighbors may not be a true $R$-near neighbor, a final pruning algorithm is used to remove the false neighbors.

Despite being efficient in storage and search costs, MIH cannot be applied to the angular preserving binary codes, since it is originally designed to solve the $R$-near neighbor problem in the Hamming space. In the rest of this section, we propose our {\it Angular Multi-index Hashing} (AMIH) technique for fast and exact search among angular preserving binary codes which addresses our second research question (RQ2).

Instead of populating one large hash table with binary codes, AMIH creates multiple smaller hash tables. To populate such smaller hash tables, each binary code $\mathbf{b}\in \{0,1\}^p$ is partitioned into $m$ disjoint substrings $\mathbf{b}^{(1)},\ldots,\mathbf{b}^{(m)}$. For the sake of simplicity, in the following, we assume that $p$ is divisible by $m$ and use the notation $w=\frac{p}{m}$. As a result, the $s$-th hash table, $s\in\{1,\ldots,w\}$, is populated with $\mathbf{b}_i^{(s)}$ $i\in\{1,\ldots,n\}$. To retrieve the $(r_1,r_2)$-near neighbors of the query, $\mathbf{q}$ is similarly partitioned into $m$ substrings, $\mathbf{q}^{(1)},\ldots,\mathbf{q}^{(m)}$. 

The following proposition establishes the relationship between the Hamming distance tuple of two binary codes and their substrings.

{\it Proposition 4: If $\mathbf{b}$ lies at a Hamming distance tuple of at most $(r_1,r_2)$ from $\mathbf{q}$, then:
\begin{equation}
\begin{split}
\exists\quad  0<t\leq m\quad  s.t. &\quad \lVert\mathbf{q}^{(t)}-\mathbf{b}^{(t)}\rVert_H\leq\floor{\frac{r_1+r_2}{m}} \\
								& \land \quad	r_{1\rightarrow 0}^{\mathbf{q}^{(t)},\mathbf{b}^{(t)}_i}\leq r_1 \\
								&  \land \quad r_{0\rightarrow 1}^{\mathbf{q}^{(t)},\mathbf{b}^{(t)}_i}\leq r_2.
\end{split}
\end{equation}

}

The first condition follows from the Pigeon-hole principle. If in all of the $m$ substrings, the Hamming distance is strictly greater than $\floor{\frac{r_1+r_2}{m}}$, then we have $\lVert\mathbf{q}-\mathbf{b}\rVert_H\geq m(\floor{\frac{r_1+r_2}{m}}+1)$. This contradicts the assumption that $\mathbf{b}$ lies at a Hamming distance of at most $r_1+r_2$ from $\mathbf{q}$. The second and the third conditions must in fact hold for all substrings, because if we have $r_{1\rightarrow 0}^{\mathbf{q}^{(t)},\mathbf{b}^{(t)}_i}> r_1$, then we should have $r_{1\rightarrow 0}^{\mathbf{q},\mathbf{b}_i}>r_1$. Similarly, if we have $r_{0\rightarrow 1}^{\mathbf{q}^{(t)},\mathbf{b}^{(t)}_i}> r_2$, then we should have $r_{0\rightarrow 1}^{\mathbf{q},\mathbf{b}_i}>r_2$. Thus, $\mathbf{b}$ is not a $(r_1,r_2)$-near neighbor of $\mathbf{q}$.$\blacksquare$

In simple terms, Proposition 4 states that, if $\mathbf{b}$ is a $(r_1,r_2)$-near neighbor of $\mathbf{q}$, then at least in one of its substrings such as $t$, $\mathbf{b}^{(t)}$ must be a $(r'_1,r'_2)$-near neighbor of $\mathbf{q}^{(t)}$, where $r_1'+r_2'\leq \floor{\frac{r_1+r_2}{m}}$, $r'_1\leq r_1$ and $r'_2\leq r_2$. 

\subsection{$(r_1,r_2)$-near Neighbor Search Using Multi-index Hashing}
We have thus far established the necessary condition that facilitates the search among substrings. At the query phase, to solve a $(r_1,r_2)$-near neighbor search, AMIH first generates the tuples that satisfy the conditions of the Proposition 4. That is, to solve the $(r_1,r_2)$-near neighbor problem, AMIH generates the set of all tuples $(r'_1,r_2')$ such that $r'_1+r'_2\leq \floor{\frac{r_1+r_2}{m}}$, where  $r'_1\leq r_1$ and $r'_2\leq r_2$. This set is denoted by $\mathcal{T}_{r_1,r_2,m}$. 

\begin{example}
\normalfont Suppose $m=2$ and we are interested in solving $(3,8)$-near neighbor problem. According to Proposition 4, we need to search among tuples with a Hamming distance of at most $5=\floor{\frac{3+8}{2}}$ that satisfy the conditions in Proposition 4. These tuples are shown in Fig.~\ref{fig::tuplestocheck}. Notice that, for each tuple, the algorithm should probe all corresponding buckets in each of the hash tables.
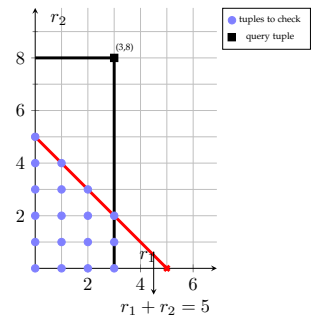
\begin{figure}[b]
\centering
\begin{tikzpicture}[scale=0.7,x=1cm,y=1cm]
	\begin{axis}[%
	clip=false,
	xmin=0,xmax=6.9,
	ymin=0,ymax=9.9,
	grid=both,
	minor tick num=1,
	axis lines = middle,
	xlabel={$r_1$},
	ylabel={$r_2$},
	y label style={at={(0.05,1)}},
	legend entries={\tiny tuples to check,\tiny query tuple},
	legend pos=outer north east,
	y=0.5cm,
	x=0.5cm,scatter/classes={%
		a={mark=*,blue!50},%
		b={mark=triangle*,brown},%
		c={mark=square*,draw=black}}]
		\addlegendimage{only marks,mark=*,color=blue!50}
		\addlegendimage{only marks,mark=square*,color=black}
\addplot[color=red,mark=x, ultra thick] coordinates {
(0,5)
(5,0)
};
\addplot[color=black,ultra thick] coordinates {
(3,0)
(3,8)
};
\addplot[color=black, ultra thick] coordinates {
(0,8)
(3,8)
};
\node at (axis cs:3.4,8.4) {\tiny (3,8)};

\addplot[scatter,only marks,%
		scatter src=explicit symbolic]%
	table[meta=label] {
x     y      label
0 0 a
0 1 a
1 0 a
0 2 a
1 1 a
2 0 a
3 0 a
2 1 a
1 2 a
0 3 a
3 1 a
2 2 a
1 3 a
0 4 a
3 2 a
2 3 a
1 4 a
0 5 a
3 8 c
};
\draw[->,>=stealth] (axis cs:4.5,0.5) to [out=90,in=90] (axis cs:4.5,-1);
\node[anchor=west] (source) at (axis cs:3, -1.5){$r_1+r_2=5$};
\end{axis}
\end{tikzpicture}
\caption{The tuples that must be checked for solving the $(3,8)$-near neighbor problem with 2 hash tables.}
\label{fig::tuplestocheck}
\end{figure}

\end{example}

Next, for each tuple such as $t=(r'_1,r'_2)$ in $\mathcal{T}_{r_1,r_2,m}$ and for each substring $\mathbf{q}^{(s)}$, $s\in\{1,\ldots,m\}$, AMIH solves the $(r'_1,r'_2)$-near neighbor problem for the query $\mathbf{q}^{(s)}$ in the $s$-th hash table. This step results in a set of candidate binary codes, denoted by $\mathcal{O}_{j,t}$. According to Proposition 4, the set $\mathcal{O}=\bigcup_{j,t}\mathcal{O}_{j,t}$ is the superset of $(r_1,r_2)$-near neighbors of $\mathbf{q}$. Finally, AMIH computes the Hamming distance tuples between $\mathbf{q}$ and all candidates in $\mathcal{O}$, discarding the tuples that are not the true $(r_1,r_2)$-near neighbors of $\mathbf{q}$.

The intuition behind this approach is that, since the number of buckets that lie at the Hamming distance tuple $(a,b)$ grows near-exponentially with the values of $a$ and $b$, it is computationally advantageous to solve multiple instances of $(a',b')$-near neighbor problem with $a'<a$ and $b'<b$, instead of solving one instance of $(a,b)$-near neighbor problem where $a$ and/or $b$ are relatively large. This requires a significantly smaller number of probings as compared to the case of deploying a single large hash table.

\subsection{Cost Analysis}

The cost analysis directly follows the performance analysis of MIH in~\cite{norouzi2014fast}. As suggested in~\cite{norouzi2014fast}, we assume that $\floor{\frac{p}{\log_2 n}}\leq m \leq \ceil{\frac{p}{\log_2 n}}$. Using AMIH, the total cost per query consists of the number of buckets that should be checked to form the candidate set $\mathcal{O}$, plus the cost of computing the Hamming distance tuple between retrieved binary codes in $\mathcal{O}$ and $\mathbf{q}$. 

We start by providing an upper bound on the number of buckets that should be checked. Since the algorithm probes identical buckets in each hash table, the number of probings equals the product of $m$ and the number of probings in a hash table.

To solve the $(r_1,r_2)$-near neighbor problem, for each tuple such as $(a,b)$ in $\mathcal{T}_{r_1,r_2,m}$, the algorithm probes the buckets that correspond to $(a,b)$. It is clear that, in the $i$-th hash table ($1\leq i \leq m$), all binary codes corresponding to the tuples in the set $\mathcal{T}_{r_1,r_2,m}$ lie at a Hammming distance of at most $\floor{\frac{r_1+r_2}{m}}$ from the $\mathbf{q}^{(i)}$ (Proposition 4). Therefore, in the $i$-th hash table, the indices of buckets that must be probed are a subset $\mathcal{C}(\mathbf{q}^{(i)},\floor{\frac{r_1+r_2}{m}})$, and we can write:
\begin{equation}
\begin{split}
\#\text{probings} &\leq \sum_{i=1}^m |\mathcal{C}(\mathbf{q}^{(i)},\floor{\frac{r_1+r_2}{m}})| \\
				& = m \times \sum_{j=0}^{\floor{\frac{r_1+r_2}{m}}} \binom{w}{j} \\
				& = m \times \sum_{j=0}^{\floor{\frac{w(r_1+r_2)}{p}}} \binom{w}{j}. \\			
\end{split}
\label{eq::cost1}
\end{equation}

Assuming that $\frac{r_1+r_2}{p}\leq 1/2$, we can use the following bound on the sum of the binomial coefficients~\cite{flum2006parameterized}.

For any $n\geq 1$ and $0<\alpha\leq 1/2$, we have:
\begin{equation}
\sum_{i=0}^{\floor{\alpha n}}\binom{n}{i} \leq 2^{H(\alpha)n}.
\end{equation}
where $H(\alpha):=-\alpha \log(\alpha)-(1-\alpha)\log(1-\alpha)$ is the binary entropy of $\alpha$.

Therefore, we can write:

\begin{equation}
\#\text{probings}\leq m\sum_{j=0}^{\floor{\frac{w(r_1+r_2)}{p}}} \binom{w}{j} \leq m2^{wH(\frac{r_1+r_2}{p})}.
\label{eq::upperbound}
\end{equation}
If binary codes are uniformly distributed in the Hamming space, the expected number of items per buckets of each hash table is $\frac{n}{2^w}$. Therefore, the expected number of items in the set $\mathcal{O}$ is:
\begin{equation}
E(|\mathcal{O}|)=\frac{n}{2^w}m\times 2^{wH(\frac{r_1+r_2}{p})}.
\label{eq::cost2}
\end{equation}
Empirically, we observed that the cost of bucket lookup is marginally smaller than the cost of verifying a candidate. If we have: {\it single lookup cost= $t\times$ single candidate test cost}, for some $t\leq 1$, then using~(\ref{eq::upperbound}) and~(\ref{eq::cost2}), we can write the total cost as:

\begin{equation}
cost \leq m2^{wH(\frac{r_1+r_2}{p})}(t+n/2^{w}).
\label{eq:totalcost}
\end{equation}
For $m\approx p/\log_2n$, by substituting $\log_2 n$ for $w$, we have:

\begin{equation}
cost = O(\frac{p}{\log_2 n}n^{H(\frac{r_1+r_2}{p})}).
\end{equation}

For reasonably small values of $\frac{r_1+r_2}{p}$, the cost is sublinear in $n$. For example, for $\frac{r_1+r_2}{p}\leq 0.1$, the expected query cost would be $O(p\sqrt{n}/\log n)$.

The space complexity of AMIH comprises: a) the cost of storing $n$ binary codes each with $p$ bits, which takes $O(np)$, and b) the cost of storing $n$ pointers to dataset items in each hash table. Each pointer can be represented in $O(\log_2n)$ bits, therefore, the cost of storing pointers would be $O(mn\log_2 n)$. For $m=\ceil{\frac{p}{\log_2 n}}$, the total storage cost is $O(np+n\log_2n)$.

\section{Experiment and Results}

In this section, we experimentally evaluate the performance of our proposed algorithm to answer the third research question (RQ3).

All techniques used in this section are implemented in C++ and compiled with identical flags. AMIH is coded on top of the MIH implementation provided by the authors of~\cite{norouzi2014fast} (all codes are compiled with GCC 4.8.4). Our implementation is publicly available at~\path{github.com/sepehr3pehr/AMIH}. The experiments have been executed on a single core of 2.0 GHz Xeon CPU with 256 Gigabytes of RAM.
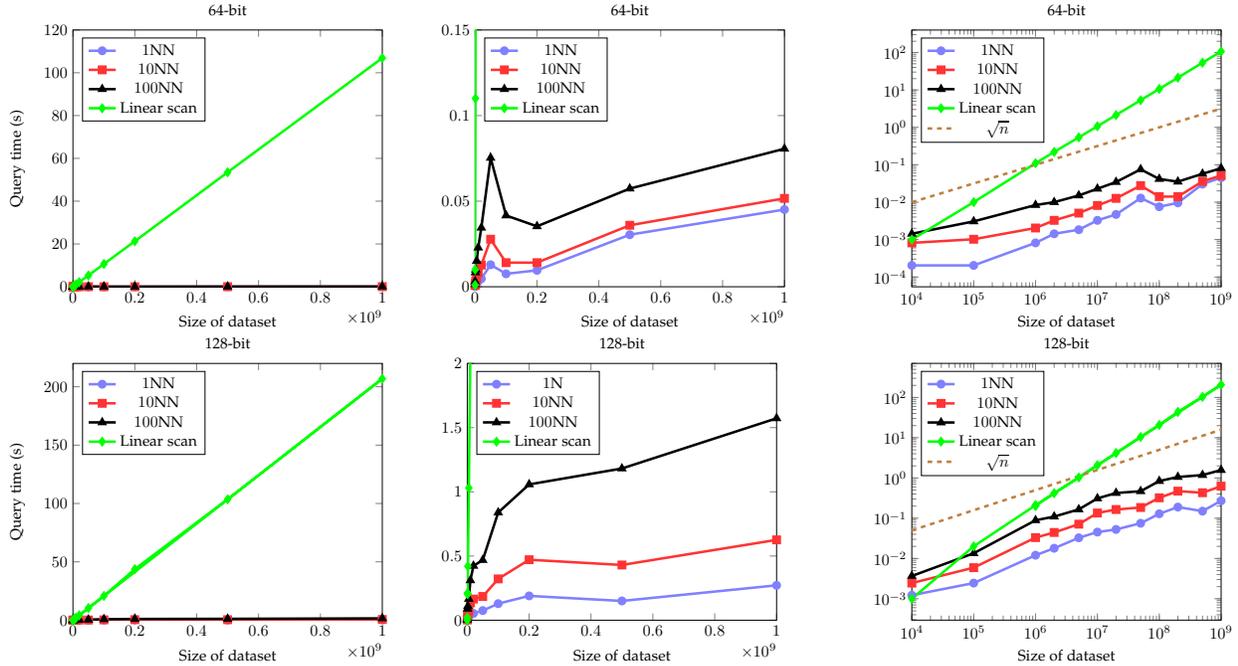
\begin{figure*}
\centering
\begin{subfigure}{.3\linewidth}
\begin{tikzpicture}[scale=0.6]
\begin{axis}[/pgfplots/tick scale binop=\times,
title=64-bit,
legend pos=north west,
ylabel={Query time (s)},
xlabel={Size of dataset},
ymin=0,ymax=120,
xmin=0,xmax=1000000000]
\addplot [select coords between index={0}{11},color=blue!50,mark=*,ultra thick] table [x=size,y=time, col sep=comma] {mih64.csv};
\addlegendentry{$1$NN}
\addplot [select coords between index={12}{23},color=red!80,mark=square*,ultra thick] table [x=size,y=time, col sep=comma] {mih64.csv};
\addlegendentry{$10$NN}
\addplot [select coords between index={24}{35},color=black,mark=triangle*,ultra thick] table [x=size,y=time, col sep=comma] {mih64.csv};
\addlegendentry{$100$NN}
\addplot [color=green,mark=diamond*,ultra thick] table [x=size,y=time, col sep=comma] {line64.csv};
\addlegendentry{Linear scan}
\end{axis}
\end{tikzpicture}
\end{subfigure}
~
\begin{subfigure}{.3\linewidth}
\begin{tikzpicture}[scale=0.6]
\begin{axis}[/pgfplots/tick scale binop=\times,
title=64-bit,
legend pos=north west,
xlabel={Size of dataset},
restrict y to domain=0:2,
ymin=0,ymax=0.15,
ytick={0,0.05,0.1,0.15},
yticklabels={$0$,$0.05$,$0.1$,$0.15$},
xmin=0,xmax=1000000000]
\addplot [select coords between index={0}{11},color=blue!50,mark=*,ultra thick] table [x=size,y=time, col sep=comma] {mih64.csv};
\addlegendentry{$1$NN}
\addplot [select coords between index={12}{23},color=red!80,mark=square*,ultra thick] table [x=size,y=time, col sep=comma] {mih64.csv};
\addlegendentry{$10$NN}
\addplot [select coords between index={24}{35},color=black,mark=triangle*,ultra thick] table [x=size,y=time, col sep=comma] {mih64.csv};
\addlegendentry{$100$NN}
\addplot [color=green,mark=diamond*,ultra thick] table [x=size,y=time, col sep=comma] {line64.csv};
\addlegendentry{Linear scan}
\end{axis}
\end{tikzpicture}
\end{subfigure}
~
\begin{subfigure}{.3\linewidth}
\begin{tikzpicture}[scale=0.6]
\begin{axis}[
title=64-bit,
legend pos=north west,
xmode=log,
ymode=log,
xlabel={Size of dataset},
xmin=10000,xmax=1000000000]
\addplot [select coords between index={0}{11},color=blue!50,mark=*,ultra thick] table [x=size,y=time, col sep=comma] {mih64.csv};
\addlegendentry{$1$NN}
\addplot [select coords between index={12}{23},color=red!80,mark=square*,ultra thick] table [x=size,y=time, col sep=comma] {mih64.csv};
\addlegendentry{$10$NN}
\addplot [select coords between index={24}{35},color=black,mark=triangle*,ultra thick] table [x=size,y=time, col sep=comma] {mih64.csv};
\addlegendentry{$100$NN}
\addplot [color=green,mark=diamond*,ultra thick] table [x=size,y=time, col sep=comma] {line64.csv};
\addlegendentry{Linear scan}
\addplot [color=brown,ultra thick,dashed,mark=none] table [x=n,y expr=\thisrowno{1}*0.0001, col sep=comma] {sqs.csv};
\addlegendentry{$\sqrt{n}$}
\end{axis}
\end{tikzpicture}
\end{subfigure}

\begin{subfigure}{.3\linewidth}
\begin{tikzpicture}[scale=0.6]
\begin{axis}[/pgfplots/tick scale binop=\times,
title=128-bit,
legend pos=north west,
xlabel={Size of dataset},
ylabel={Query time (s)},
ymin=0,ymax=220,
xmin=0,xmax=1000000000]
\addplot [select coords between index={0}{11},color=blue!50,mark=*,ultra thick] table [x=size,y=time, col sep=comma] {mih128.csv};
\addlegendentry{$1$NN}
\addplot [select coords between index={12}{23},color=red!80,mark=square*,ultra thick] table [x=size,y=time, col sep=comma] {mih128.csv};
\addlegendentry{$10$NN}
\addplot [select coords between index={24}{35},color=black,mark=triangle*,ultra thick] table [x=size,y=time, col sep=comma] {mih128.csv};
\addlegendentry{$100$NN}
\addplot [color=green,mark=diamond*,ultra thick] table [x=size,y=time, col sep=comma] {line128.csv};
\addlegendentry{Linear scan}
\end{axis}
\end{tikzpicture}
\end{subfigure}
~
\begin{subfigure}{.3\linewidth}
\begin{tikzpicture}[scale=0.6]
\begin{axis}[/pgfplots/tick scale binop=\times,
title=128-bit,
legend pos=north west,
xlabel={Size of dataset},
ymin=0,ymax=2,
restrict y to domain=0:10,
ytick={0,0.5,1,1.5,2},
xmin=0,xmax=1000000000]
\addplot [select coords between index={0}{11},color=blue!50,mark=*,ultra thick] table [x=size,y=time, col sep=comma] {mih128.csv};
\addlegendentry{$1$N}
\addplot [select coords between index={12}{23},color=red!80,mark=square*,ultra thick] table [x=size,y=time, col sep=comma] {mih128.csv};
\addlegendentry{$10$NN}
\addplot [select coords between index={24}{35},color=black,mark=triangle*,ultra thick] table [x=size,y=time, col sep=comma] {mih128.csv};
\addlegendentry{$100$NN}
\addplot [color=green,mark=diamond*,ultra thick] table [x=size,y=time, col sep=comma] {line128.csv};
\addlegendentry{Linear scan}
\end{axis}
\end{tikzpicture}
\end{subfigure}
~
\begin{subfigure}{.3\linewidth}
\begin{tikzpicture}[scale=0.6]
\begin{axis}[/pgfplots/tick scale binop=\times,
title=128-bit,
legend pos=north west,
xmode=log,
ymode=log,
xlabel={Size of dataset},
xmin=10000,xmax=1000000000]
\addplot [select coords between index={0}{11},color=blue!50,mark=*,ultra thick] table [x=size,y=time, col sep=comma] {mih128.csv};
\addlegendentry{$1$NN}
\addplot [select coords between index={12}{23},color=red!80,mark=square*,ultra thick] table [x=size,y=time, col sep=comma] {mih128.csv};
\addlegendentry{$10$NN}
\addplot [select coords between index={24}{35},color=black,mark=triangle*,ultra thick] table [x=size,y=time, col sep=comma] {mih128.csv};
\addlegendentry{$100$NN}
\addplot [color=green,mark=diamond*,ultra thick] table [x=size,y=time, col sep=comma] {line128.csv};
\addlegendentry{Linear scan}
\addplot [color=brown,ultra thick,dashed,mark=none] table [x=n,y expr=\thisrowno{1}*0.0005, col sep=comma] {sqs.csv};
\addlegendentry{$\sqrt{n}$}
\end{axis}
\end{tikzpicture}
\end{subfigure}
\caption{Average search time for 64-bit and 128-bit binary codes of the SIFT dataset. AMIH and linear scan are executed to solve the $K$NN problem with $K\in$ \{1, 10, 100\}.}
\label{fig::querytime}
\end{figure*}

\subsection{Datasets}

In our experiments, we have used two non-synthetic datasets:

\noindent {\bf SIFT:} The ANN$\_$SIFT1B dataset~\cite{localitycomparison} consists of SIFT descriptors. The available dataset has been originally partitioned into $10^9$ items as the \emph{base set}, $10^4$ items as the \emph{query set}, and $10^8$ items as the \emph{learning set}. Each data item is a 128-dimensional SIFT vector.

\noindent {\bf TRC2:} The TRC2 ({\it Thomas Reuters Text Research Collection 2}) consists of 1,800,370 news stories covering a period of one year. We have used $5\times 10^5$ news as the learning set, $10^6 $ news as the base set, and the remaining as the query set. We have preprocessed the data by removing common stop words, stemming, and then considering only the 2000 most frequent words in the learning set. Thus, each news story is represented as a vector composed of 2000 word-counts.

Since the items of these datasets lie in real space, we incorporate a binary hashing technique to map the items to binary codes. For our experiments, we have used the angular-preserving mapping method called {\it Angular Quantization-based Binary Codes} (AQBC) proposed in~\cite{Gong_angular:NIPS2012_4831}, to create the dataset of binary codes. We implemented AQBC in Python following the initialization and parameter setting described in~\cite{Gong_angular:NIPS2012_4831}. We have also made our implementation of AQBC publicly available at~\path{github.com/sepehr3pehr/AQBC}.
For each dataset, the learning set is used to optimize the parameters of the hash function. Once learning is completed, the learning set is removed and the learned hash function is applied to the base and the query sets. The base set is used to populate the hash tables. Then, the angular $K$NN problem is solved for all queries points and the average performance is reported.

\subsection{AMIH vs Linear Scan}
\label{sec::amihvslinearscan}

Our first experiment compares the performance of linear scan with AMIH in terms of the search speed. The norm of any binary code with $p$ bits ranges from 0 to $\sqrt{p}$. Thus, to increase the speed of the linear scan, we initialize a look up table with all the possible norm values. Moreover, as the term $\sqrt{\lVert\mathbf{q}\rVert_1}$ in the denominator of~(\ref{eq:cosine_r1r2}) is independent of $\mathbf{b}_i$, there is no need to account for its value in searching.

 We observed that the performance of linear scan is virtually independent of $K$ (number of nearest neighbors). Consequently, for the sake of comparison, in the following, we only use the result of the linear scan for the 1NN problem. Note that the linear scan can benefit from caching, as it performs sequential memory access. Otherwise, it would be much slower.

\def\arraystretch{1.5}
\begin{table}
    \centering
	\footnotesize
            \caption{Speedup gains that AMIH achieves in comparison to linear scan. The last line shows the average query time of linear scan in seconds.}
        \begin{tabular}{c c c c c c }
            \toprule
            \midrule
              &   & \multicolumn{2}{c} {\textbf{SIFT 1B}} & \multicolumn{2}{c}{\textbf{TRC2}} \\ \cmidrule{2-6}
               & \textbf{$\#$ bits:}    & \textbf{64}  & \textbf{128} & \textbf{64} & \textbf{128}  \\ \cmidrule{2-6}
                \multirow{3}{*}{\textbf{Speedup gain}} & 1NN & 2672 & 1035 &   106 & 7.5 \\
                	& 10NN & 2137 & 345 & 27.5 & 3.21 \\
                	& 100NN & 1336 & 138 & 9.1 & 2.1\\ \cmidrule{2-6}
                	& Linear scan (s): & 106 & 207 & 0.110 & 0.206 \\
              
            \midrule
            \bottomrule
        \end{tabular}
        \centering
        \label{tab::speedup}
\end{table}

Fig.~\ref{fig::querytime} shows the average query time as a function of the dataset size for 64-bit and 128-bit binary codes. In all experiments, the value of $m$ (number of hash tables) for AMIH is set to $\frac{p}{\log_2 n}$, following~\cite{aizawa1pqtable,greene1994multi,norouzi2014fast}. The leftmost graphs show the search time in seconds in terms of the data base size. It is apparent that AMIH is significantly faster than the linear scan for a broad range of dataset sizes and $K$ values. To differentiate between the performance of AMIH for different values of $K$, the middle graphs show the zoomed version of the leftmost graphs, and the rightmost graphs are plotted using logarithmic scale. As Figs.~\ref{fig::querytime} illustrates, for linear scan, the query time grows linearly with the dataset size, whereas the query time of AMIH increases with the square root of the size. Consequently, the difference between the query times of the two techniques is more significant for larger datasets. For instance, linear scan spends more than three minutes to report the nearest neighbor in the $10^9$ SIFT dataset with 128-bit codes, while AMIH takes about a quarter of a second. The dashed line on log-log plots shows the growth rate of the $\sqrt{n}$ up to a constant factor. The evident similarity between the slope of this function and that of AMIH query time indicates that, even for non-uniform distributions, AMIH can achieve sublinear search time.

\begin{figure}
\centering
\begin{subfigure}{.5\linewidth}
\begin{tikzpicture}[scale=0.55]
\begin{axis}[ 
title=64-bit,
	legend pos=south west,
	ymin=10,ymax=100,
	xmode=log,
    xlabel={Size of dataset},
    ylabel={Percentage of queries}]]
\addplot table [x=a, y=b, col sep=comma] {1NN64.csv};
\addplot table [x=a, y=c, col sep=comma] {10NN64.csv};
\addplot [mark=triangle*, mark options={scale=1.5}] table [x=a, y=c, col sep=comma] {100NN64.csv};
\legend{1NN,10NN,100NN}
\end{axis}
\end{tikzpicture}
\end{subfigure}~
\begin{subfigure}{.5\linewidth}
\begin{tikzpicture}[scale=0.55]
\begin{axis}[
	title=128-bit,
	legend pos=south west,
	ymin=60,ymax=100,
	xmode=log,
    xlabel={Size of dataset},
    ]]
\addplot table [x=a, y=c, col sep=comma] {1NN128.csv};
\addplot table [x=a, y=c, col sep=comma] {10NN128.csv};
\addplot [mark=triangle*, mark options={scale=1.5}] table [x=a, y=c, col sep=comma] {100NN128.csv};
\legend{1NN,10NN,100NN}
\end{axis}
\end{tikzpicture}
\end{subfigure}
\caption{The percentage of queries for which  the required radius of search gets larger than $\hat{r}$.}
\label{fig::secondcase}
\end{figure}
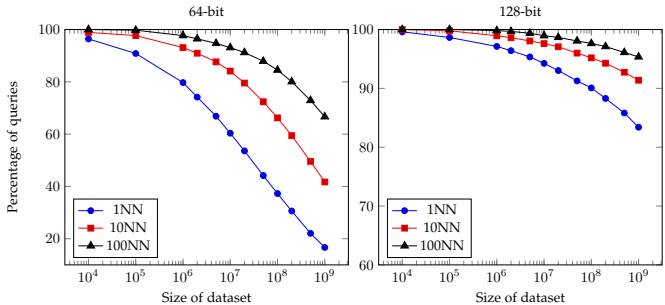

Fig.~\ref{fig::secondcase} shows the percentage of queries for which  the required radius of search gets larger than $\hat{r}$. As the size of the dataset grows, the number of empty buckets reduces, and the algorithm finds the nearest neighbors within a smaller search radius. Similarly, for shorter binary codes, the number of buckets reduces, and in turn,  AMIH retrieves items before the search radius reaches $\hat{r}$.

Table~\ref{tab::speedup} includes the speed up factors achieved by AMIH versus linear scan. Each entry in the table indicates the average query time using linear scan over the average query time using AMIH for a specific value of $K$. Interestingly, AMIH solves the angular $K$NN problem up to hundreds and even thousands of times faster than linear scan. In particular, AMIH can solve the 100NN problem 138 times faster than the linear scan on a dataset of $10^9$ binary codes each with 128 bits.

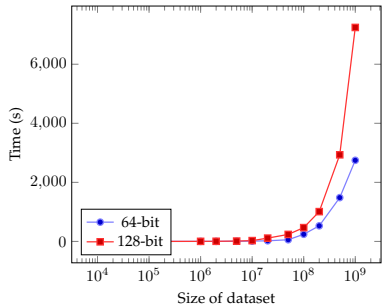
\begin{figure}
\centering
\begin{tikzpicture}[scale=0.6]
\begin{axis}[
xmode=log,
xlabel={Size of dataset},
ylabel={Time (s)},
legend pos=south west
]									
\addplot+[color=blue!50,thick] table [x=size,y=time, col sep=comma]{indexingtime64.csv};
\addplot+[color=red!80,thick] table [x=size,y=time, col sep=comma]{indexingtime128.csv};
\legend{64-bit, 128-bit}
\end{axis}
\end{tikzpicture}

\caption{Indexing time of AMIH on $10^9$ SIFT dataset.}
\label{fig::indexingtime}
\end{figure}

While the linear scan technique does not rely on any indexing phase, AMIH requires each binary code to be indexed in $m$ hash tables. The indexing time for AMIH using the SIFT dataset is shown in Fig.~\ref{fig::indexingtime}. For 64 and 128 bit codes, the indexing phase takes about 1 and 2 hours, respectively.

\subsection{AMIH vs Approximate Techniques}
Due to the curse of dimensionality, linear scan is theoretically the fastest exact technique for solving the angular $K$NN problem in its general setting. However, a handful of approximation algorithms exist that provide sublinear search time for this problem. The most well-known representative among these is~\emph{Locality Sensitive Hashing} (LSH)~\cite{Indyk:1998:ANN:276698.276876}, which offers a (provably) sublinear search time. In addition to LSH, some applied approximation algorithms have been proposed which work promising in realworld applications, such as \emph{KGraph}~\cite{kgraph} and \emph{Annoy}~\cite{annoy}, but do not necessarily guarantee (efficient) worst case analysis. In this section, we compare AMIH with some of the well-known approximation techniques for the task of nearest neighbor search ($1$NN). 

The comparison between AMIH and other approximation techniques is performed in two different scenarios:
\begin{itemize}

\item First, we investigate the performance of different techniques for solving the angular nearest neighbor problem in the binary space. Similar to the experiments of section~\ref{sec::amihvslinearscan}, we assume that we are given a binary dataset and the goal entails solving the angular nearest neighbor search for binary query points.

\item In the second scenario, we assume that the original dataset lies in the real space. Given the dataset, we apply the approximate algorithms to the real dataset. However, since AMIH can be only applied to binary codes, we first use a hashing algorithm to map the dataset to binary codes and then use AMIH to solve the $K$ nearest neighbor search within the binary dataset. Finally, among the $K$ retrieved points, we select the one that is the closet to the query in the real space. Therefore, the returned nearest neighbor in this setting is approximate with respect to the original data points lying in the real space. This scenario mainly targets applications in which the original dataset items do not lie in the binary space because for binary datasets the result of AMIH is exact.
\end{itemize}

\subsubsection{Approximate Techniques Used for Comparison}
For both scenarios, we compare AMIH with three state-of-the-art ANN techniques. Discussing the details of these techniques is beyond the scope of this paper but we briefly introduce each of them here.

\textbf{Crosspolytope LSH}~\cite{andoni2015practical} is a recently proposed LSH technqiue for solving the angular nearest neighbor search problem. The general idea behind LSH is to randomly partition the feature space using a specific family of hash functions that map similar items into the same buckets with high probability. Given such hash functions, during the preprocessing step, all items of the dataset are inserted in to $l$ hash tables corresponding to $l$ randomly chosen hash functions (each hash function represents a partitioning of the space). To find the nearest neighbors the query vector is similarly mapped $l$ times, and the items in the corresponding $l$ hash buckets are retrieved as the candidates for the nearest neighbor. The algorithm then passes through the retrieved points to find the closest one to the query. This variant of LSH is often called~\emph{Single-Probe} (SP) LSH as it probes only one bucket per hash table.

We also compare the performance of AMIH with the \emph{MultiProbe} (MP) variant of the crosspolytope LSH. Multiprobe LSH~\cite{lv2007multi} is an extension of LSH that can achieve significant space reduction via reducing the number of required hash tables. The basic idea of multiprobe LSH is to not only consider the main bucket, where the query falls, but also probe other buckets that are~\emph{close} to the main bucket in every hash table. For our comparisons, we use the multiprobe variant of the crosspolytope LSH described in~\cite{andoni2015practical}. The source code of this method has been made publicly available as a part of the \emph{FAst Lookups of Cosine and Other Nearest Neighbors} (FALCONN) library~\cite{falconn}.

\textbf{KGraph}~\cite{kgraph} performs the nearest neighbor search by building a $K$NN graph over the datapoints. In the graph, each node corresponds to a data point and is connected to its $M$ nearest point where $M$ needs to be tuned. During query phase, the algorithm starts from one of the nodes and follows the paths with shorter distances to find the approximate nearest neighbors.

\textbf{Annoy}~\cite{annoy} decomposes the search space using multiple trees to achieve sublinear search time. At each non-leaf node, a random hyperplane is formed by taking the equidistant hyperplane of two randomly selected data points. Each internal node therefore divides the space into two subspaces where each subspace contains at least one data point. Each leaf node contains a subset of datapoints that lie in the region of space defined by the leaf's ancestors. To find the nearest neighbor, the search algorithm only considers the subspaces where the query fall in. Annoy incorporates a forest of such trees to increase the probability of collision between query and its nearest neighbor in at least one leaf node.

\subsubsection{Experimental setting}
In all experiments, for single-probe crosspolytope (SP-CP) LSH, KGraph and Annoy, we use the parameter settings of \emph{ann-benchmark}~\cite{aumuller2017ann} which is a tool of standardizing benchmarking for approximate nearest neighbor search algorithms. The SP-CP setting in ann-benchmark incorporates a fixed value for the number of hash functions per hash tables, $k=16$, and let $l$ vary from 1 to 1416. For multiprobe crosspolytope (MP-CP) LSH (which is absent in ann-benchmark), we follow the parameter setting of~\cite{andoni2015practical}. In particular, for MP-CP, we use only 10 hash tables ($l=10$) in each experiment. As stated in~\cite{andoni2015practical}, the goal of this choice is to keep the additional memory occupied by LSH comparable to the amount of memory needed for storing dataset. This is perhaps the most practical and interesting scheme since large memory overheads are impossible for massive datasets. To set $k$ (number of hash functions per hash table), we try different values for this parameter and select the one with the minimum query time. To do that, following~\cite{falconn}, for each value of $k~\in\{10\ldots, 30\}$, we use binary search to find the minimum number of probes that results in a near-perfect recall rate ($\geq 0.9$), meaning that 10 percent of returned neighbors are not exact. After fixing $k$ and $l$, the number of probes per hash table is gradually increased (which results in higher recall rates) and for each value the average query time is reported.

In the following, we compare the performance of different techniques in terms of the average query time and the memory requirement. Note that the memory cost reported here is the additional memory required by each technique to build its data structure (memory required to store the raw dataset is not included). The experiments of this section are executed on a single core 3.0 GHz CPU with 32 GB of memory.

Before discussing the results, we would like to note that the ann-benchmark basically is not designed for scenarios with low memory budget. We observed that the settings used for techniques such as KGraph and Annoy require amount of memory that is much large than the memory required to the store the dataset. Also for LSH, the benchmark only uses single-probe LSH. The main reason for this choice is that, in comparison to multiprobe LSH, single-probe LSH achieves better query time when RAM budget is not a matter of concern. The memory cost of ann-benchmark is perhaps the main reason why larger datasets such as 1 billion SIFT vectors (that we used in the first experiments) are absent in the benchmark (all datasets in ann-benchmark have around 1 million points). The authors of~\cite{andoni2015practical} have also explicitly mentioned that the experiments of ann-benchmark are not efficient for low RAM budget scenarios~\cite{falconn}.

\subsubsection{Nearest Neighbor Search in Binary Space}
Here we use the ANN$\_$SIFT1M~\cite{product} dataset which consists of 1 million 128D SIFT vectors for the base set and 10000 query items. Similar to section~\ref{sec::amihvslinearscan}, the dataset is binarized to 64-bit and 128-bit codes by applying the AQBC technique. The binary dataset is then fed to each technique and the average query time as well as memory cost is reported.

\begin{figure*}
\begin{subfigure}{.4\linewidth}
\begin{tikzpicture}[scale=0.7]

\begin{axis}[%
width=4.5in,
height=2.5in,
xmin=0,
xmax=1,
xlabel={Recall},
ymode=log,
ymin=0,
ymax=1,
ylabel={Query time (s)},
xmajorgrids,
ymajorgrids,
title={64-bit},
legend style={legend cell align=left, align=left, draw=white!15!black, legend pos=north west}
]
\addplot+[ultra thick]
  table[row sep=crcr]{%
0.0098	4.80966329574585e-05\\
0.0184	6.15109443664551e-05\\
0.0277	7.51693487167358e-05\\
0.0366	8.74226331710815e-05\\
0.0451	0.000100767397880554\\
0.0555	0.000114865183830261\\
0.0637	0.000127548360824585\\
0.0733	0.000160059809684753\\
0.0831	0.000153832340240479\\
0.0897	0.000164335775375366\\
0.0977	0.000179214382171631\\
0.114	0.000208557915687561\\
0.133	0.000234980201721191\\
0.1488	0.000261921286582947\\
0.1662	0.000287408065795898\\
0.1801	0.000312620782852173\\
0.2015	0.00035331084728241\\
0.2225	0.000393797302246094\\
0.2427	0.000436516189575195\\
0.262	0.000476994132995605\\
0.2899	0.000526670455932617\\
0.3153	0.000577255177497864\\
0.345	0.000649847197532654\\
0.3704	0.000711112403869629\\
0.3991	0.000798080396652222\\
0.4307	0.000870080447196961\\
0.4637	0.000966124820709229\\
0.4961	0.00105264852046967\\
0.531	0.00115830302238464\\
0.5686	0.00128406283855438\\
0.6012	0.00141149075031281\\
0.6351	0.00156873137950897\\
0.6674	0.00171981086730957\\
0.7006	0.00189364476203918\\
0.7303	0.00207481229305267\\
0.7636	0.00228241264820099\\
0.7945	0.00253658514022827\\
0.8232	0.00277260355949402\\
0.8512	0.00303072762489319\\
0.8729	0.00330991222858429\\
0.896	0.00365935733318329\\
0.9149	0.00402506742477417\\
0.9326	0.00444307720661163\\
0.9485	0.00485823385715485\\
0.9593	0.00532643311023712\\
0.9701	0.00591319205760956\\
0.9774	0.00646495933532715\\
0.9832	0.00722290525436401\\
0.9885	0.00754971132278442\\
0.9919	0.00827823913097382\\
0.9938	0.00910813679695129\\
0.9952	0.00994522800445557\\
0.9965	0.0107946822404861\\
0.9979	0.0118499357223511\\
0.9988	0.0129672160387039\\
0.9992	0.014144619846344\\
0.9997	0.015436776471138\\
0.9998	0.0168334487199783\\
};
\addlegendentry{SP-CP}

\addplot+[ultra thick]
  table[row sep=crcr]{%
0.378	0.00244561746120453\\
0.4756	0.00343479783535004\\
0.5498	0.00422133746147156\\
0.6188	0.00498235175609589\\
0.6631	0.00571852390766144\\
0.8147	0.0090133279800415\\
0.9253	0.0149491507291794\\
0.9638	0.0207534953594208\\
0.9824	0.0266391820430756\\
0.9899	0.0323277034282684\\
0.9941	0.0379437507867813\\
0.9972	0.0434341568231583\\
0.9977	0.0485798802375794\\
0.9986	0.05358822889328\\
0.9989	0.0585979530572891\\
};
\addlegendentry{KGraph}

\addplot+[ultra thick]
  table[row sep=crcr]{%
0.0379	0.000595726203918457\\
0.043	0.000782838082313538\\
0.0499	0.000830319213867188\\
0.0607	0.000670361638069153\\
0.0702	0.000841883993148804\\
0.0807	0.000951566433906555\\
0.0972	0.000872640442848206\\
0.1112	0.000919775366783142\\
0.1319	0.00133485369682312\\
0.1808	0.00135283920764923\\
0.2163	0.00150451400279999\\
0.2465	0.00206790835857391\\
0.2834	0.00237440659999847\\
0.3277	0.00265619106292725\\
0.3849	0.00334981462955475\\
0.4338	0.00452833871841431\\
0.4819	0.00474186797142029\\
0.5471	0.00578349335193634\\
0.6542	0.0101094648599625\\
0.7213	0.0103158603906631\\
0.775	0.0118153691530228\\
0.8167	0.0187966796875\\
0.8707	0.0192533293485641\\
0.9078	0.0215357322454453\\
0.9305	0.0354310845851898\\
0.9567	0.0359803482055664\\
0.9773	0.0397934345245361\\
0.9902	0.0893918765306473\\
0.9959	0.0795558308601379\\
0.9986	0.0886065073490143\\
0.9987	0.140582766819\\
0.9995	0.140596117734909\\
0.9998	0.156823402452469\\
1	0.256171918845177\\
1	0.260213068556786\\
1	0.224276238822937\\
};
\addlegendentry{Annoy}
\addplot+[ultra thick, mark=square*] table [x=recall,y=qt, col sep=comma] {mp-cp-64.csv};;
\addlegendentry{MP-CP}

\addplot[violet,only marks, mark size=4.0pt, mark=diamond*] coordinates {( 1, 0.0006)};
\addlegendentry{AMIH}
\end{axis}
\end{tikzpicture}%

\end{subfigure}
\quad\quad\quad\quad\quad
\begin{subfigure}{.4\linewidth}
\begin{tikzpicture}[scale=0.7]

\begin{axis}[%
width=4.5in,
height=2.5in,
xmin=0,
xmax=1,
xlabel={Recall},
ymode=log,
ymin=1000,
ymax=100000000,
yminorticks=true,
ylabel={Index size (KB)},
xmajorgrids,
ymajorgrids,
title={64-bit},
legend style={legend cell align=left, align=left, draw=white!15!black, legend pos=north west}
]
\addplot+[ultra thick]
  table[row sep=crcr]{%
0.0098	4484\\
0.0184	10012\\
0.0277	12552\\
0.0366	21672\\
0.0451	29872\\
0.0555	38264\\
0.0637	46656\\
0.0733	54984\\
0.0831	59404\\
0.0897	63696\\
0.0977	68116\\
0.114	76956\\
0.133	85924\\
0.1488	94632\\
0.1662	103464\\
0.1801	112296\\
0.2015	125608\\
0.2225	138868\\
0.2427	152192\\
0.262	165452\\
0.2899	183132\\
0.3153	200808\\
0.345	223016\\
0.3704	244980\\
0.3991	271624\\
0.4307	297992\\
0.4637	328992\\
0.4961	359848\\
0.531	395196\\
0.5686	435104\\
0.6012	479144\\
0.6351	527764\\
0.6674	580776\\
0.7006	638212\\
0.7303	700136\\
0.7636	770768\\
0.7945	850360\\
0.8232	934308\\
0.8512	1027032\\
0.8729	1128648\\
0.896	1243516\\
0.9149	1367240\\
0.9326	1504200\\
0.9485	1654480\\
0.9593	1822312\\
0.9701	2003468\\
0.9774	2202408\\
0.9832	2423208\\
0.9885	2666212\\
0.9919	2931304\\
0.9938	3223036\\
0.9952	3545320\\
0.9965	3898912\\
0.9979	4287848\\
0.9988	4716292\\
0.9992	5189040\\
0.9997	5705988\\
0.9998	6276072\\
};
\addlegendentry{SP-CP}

\addplot+[ultra thick]
  table[row sep=crcr]{%
0.378	4583088\\
0.4756	4583088\\
0.5498	4560048\\
0.6188	4559808\\
0.6631	4560008\\
0.8147	4559796\\
0.9253	4559908\\
0.9638	4560000\\
0.9824	4560000\\
0.9899	4559916\\
0.9941	4560000\\
0.9972	4559784\\
0.9977	4560016\\
0.9986	4559772\\
0.9989	4560008\\
};
\addlegendentry{KGraph}

\addplot+[ultra thick]
  table[row sep=crcr]{%
0.0379	1731940\\
0.043	2912152\\
0.0499	4921200\\
0.0607	1729792\\
0.0702	2912152\\
0.0807	4921200\\
0.0972	1727384\\
0.1112	2912152\\
0.1319	4921264\\
0.1808	1722160\\
0.2163	2912152\\
0.2465	4921200\\
0.2834	1723392\\
0.3277	2912152\\
0.3849	4921200\\
0.4338	1723392\\
0.4819	2912152\\
0.5471	4921200\\
0.6542	1723456\\
0.7213	2912152\\
0.775	4921196\\
0.8167	1723392\\
0.8707	2912152\\
0.9078	4921200\\
0.9305	1723392\\
0.9567	2912152\\
0.9773	4921200\\
0.9902	1723392\\
0.9959	2912152\\
0.9986	4921200\\
0.9987	1723392\\
0.9995	2912152\\
0.9998	4921200\\
1	4921200\\
1	1723392\\
1	2912152\\
};
\addlegendentry{Annoy}

\addplot+[ only marks, mark size=3.0pt, mark=square*] coordinates {( 0.9, 95044)};
\addlegendentry{MP-CP}
\addplot[violet,only marks, mark size=4.0pt, mark=diamond*] coordinates {( 1, 63000)};
\addlegendentry{AMIH}

\addplot[only marks, mark size=4.0pt, mark=oplus*, fill=yellow!80!black,draw=black] coordinates {( 0, 6800)};
\addlegendentry{Dataset}

\end{axis}
\end{tikzpicture}
\end{subfigure}

\begin{subfigure}{.4\linewidth}

\begin{tikzpicture}[scale=0.7]

\begin{axis}[%
width=4.5in,
height=2.5in,
xmin=0,
xmax=1,
xlabel={Recall},
ymode=log,
ymin=0,
ymax=1,
ylabel={Query time (s)},
xmajorgrids,
ymajorgrids,
title={128-bit},
legend style={legend cell align=left, align=left, draw=white!15!black, legend pos=north west}
]
\addplot+[ultra thick]
  table[row sep=crcr]{%
0.1676	0.000808188891410828\\
0.2659	0.00130796935558319\\
0.3434	0.00171973729133606\\
0.408	0.00204231603145599\\
0.4599	0.00244673964977264\\
0.5045	0.002848237657547\\
0.554	0.00368716371059418\\
0.5851	0.00395438265800476\\
0.6113	0.00420707573890686\\
0.6378	0.00445463926792145\\
0.665	0.00483089034557342\\
0.7155	0.00588318722248077\\
0.7525	0.00664893362522125\\
0.7838	0.00739057466983795\\
0.8071	0.00786639478206634\\
0.8318	0.00869837160110474\\
0.8555	0.00953194105625153\\
0.8812	0.0110334859132767\\
0.8978	0.0121312863826752\\
0.9142	0.0131181427717209\\
0.9272	0.0142490114212036\\
0.9406	0.0157369428873062\\
0.9531	0.0171277657270432\\
0.9625	0.0190416934728622\\
0.9705	0.0207844480276108\\
0.977	0.0226815931558609\\
0.9812	0.0251522896766663\\
0.9854	0.0279826129674912\\
0.9894	0.0300848503828049\\
0.9914	0.0320614491224289\\
0.9938	0.0343670062541962\\
0.9951	0.036245631814003\\
0.9962	0.0405420154333115\\
0.9975	0.0430484557151794\\
0.9981	0.0471383409023285\\
0.9986	0.0519062671661377\\
0.999	0.0544031291484833\\
0.999	0.0520745306253433\\
0.9993	0.0573971663475037\\
0.9995	0.0615124513149262\\
0.9996	0.0664125122308731\\
0.9997	0.0849237286329269\\
0.9997	0.0713040334463119\\
0.9997	0.0893293033599853\\
0.9997	0.0783946440935135\\
0.9997	0.0748613713026047\\
0.9997	0.0708575558185577\\
0.9998	0.102940013861656\\
0.9998	0.111050234079361\\
0.9998	0.136348883271217\\
0.9998	0.120327075242996\\
0.9998	0.0936380676984787\\
0.9998	0.122402874708176\\
0.9998	0.142696851992607\\
0.9998	0.127172226285934\\
0.9998	0.106957230377197\\
0.9998	0.0981677906990051\\
0.9998	0.115371130299568\\
};
\addlegendentry{SP-CP}

\addplot+[ultra thick]
  table[row sep=crcr]{%
0.7774	0.00350668585300446\\
0.8425	0.00399276783466339\\
0.8982	0.0039096289396286\\
0.9232	0.0041352240562439\\
0.939	0.00474294612407684\\
0.97	0.00572341775894165\\
0.9912	0.00805422120094299\\
0.9953	0.0104328295946121\\
0.996	0.0202754629611969\\
0.9965	0.0157589318037033\\
0.997	0.0124866815090179\\
0.9976	0.0174190616130829\\
0.9979	0.014097992682457\\
0.9981	0.0218650067329407\\
0.999	0.0191190493345261\\
};
\addlegendentry{KGraph}

\addplot+[ultra thick]
  table[row sep=crcr]{%
0.4133	0.00294865076541901\\
0.4369	0.00136447360515594\\
0.4631	0.00348200325965881\\
0.5309	0.000996600699424744\\
0.5611	0.00165344731807709\\
0.5843	0.00312831106185913\\
0.649	0.00121920094490051\\
0.6835	0.00197104873657227\\
0.7076	0.00391394200325012\\
0.7904	0.0020068808555603\\
0.8171	0.00276970925331116\\
0.8347	0.00520102596282959\\
0.872	0.00320309674739838\\
0.8883	0.00393061826229095\\
0.9009	0.00672418818473816\\
0.9269	0.00512537078857422\\
0.9375	0.00599667377471924\\
0.9436	0.00833814644813538\\
0.9671	0.0101354122877121\\
0.9751	0.0108253435850143\\
0.9767	0.0125457879066467\\
0.9852	0.0174107917785645\\
0.989	0.0164176001787186\\
0.9897	0.0221360235452652\\
0.9926	0.0283688384771347\\
0.9951	0.0266229722499847\\
0.9956	0.0311771394014358\\
0.9974	0.0531105849027634\\
0.9983	0.0511243770122528\\
0.9987	0.054898792719841\\
0.9989	0.0851532160043716\\
0.9991	0.0828531476736069\\
0.9992	0.0754288329839706\\
0.9994	0.134009651136398\\
0.9995	0.121919530177116\\
0.9996	0.138108175802231\\
};
\addlegendentry{Annoy}
\addplot+[ultra thick, mark=square*] table [x=recall,y=qt, col sep=comma] {mp-cp-128.csv};;
\addlegendentry{MP-CP}

\addplot[violet,only marks, mark size=4.0pt, mark=diamond*] coordinates {( 1, 0.0015)};
\addlegendentry{AMIH}
\end{axis}
\end{tikzpicture}%

\end{subfigure}
\quad\quad\quad\quad\quad
\begin{subfigure}{.4\linewidth}
\begin{tikzpicture}[scale=0.7]

\begin{axis}[%
width=4.5in,
height=2.5in,
xmin=0,
xmax=1,
xlabel={Recall},
ymode=log,
ymin=1000,
ymax=100000000,
yminorticks=true,
ylabel={Index size (KB)},
xmajorgrids,
ymajorgrids,
title={128-bit},
legend style={legend cell align=left, align=left, draw=white!15!black, legend pos=north west}
]
\addplot+[ultra thick]
  table[row sep=crcr]{%
0.1676	4484\\
0.2659	10752\\
0.3434	15864\\
0.408	21804\\
0.4599	29884\\
0.5045	38212\\
0.554	46520\\
0.5851	54852\\
0.6113	59272\\
0.6378	63692\\
0.665	68112\\
0.7155	77072\\
0.7525	86040\\
0.7838	94876\\
0.8071	103580\\
0.8318	112412\\
0.8555	125660\\
0.8812	138916\\
0.8978	152180\\
0.9142	165440\\
0.9272	183120\\
0.9406	200796\\
0.9531	222876\\
0.9625	244968\\
0.9705	271548\\
0.977	297980\\
0.9812	328980\\
0.9854	359836\\
0.9894	395184\\
0.9914	434964\\
0.9938	479132\\
0.9951	527752\\
0.9962	580764\\
0.9975	638204\\
0.9981	700124\\
0.9986	770756\\
0.999	934360\\
0.999	850284\\
0.9993	1027020\\
0.9995	1128636\\
0.9996	1243500\\
0.9997	2003456\\
0.9997	1504188\\
0.9997	2202264\\
0.9997	1822428\\
0.9997	1654404\\
0.9997	1367228\\
0.9998	2931292\\
0.9998	3545564\\
0.9998	5705976\\
0.9998	4287708\\
0.9998	2423196\\
0.9998	4716408\\
0.9998	6275932\\
0.9998	5189028\\
0.9998	3222896\\
0.9998	2666328\\
0.9998	3898900\\
};
\addlegendentry{MP-CP}

\addplot+[ultra thick]
  table[row sep=crcr]{%
0.7774	2339348\\
0.8425	2326768\\
0.8982	2303708\\
0.9232	2303656\\
0.939	2303660\\
0.97	2303648\\
0.9912	2303668\\
0.9953	4528340\\
0.996	2303420\\
0.9965	2303332\\
0.997	2303372\\
0.9976	2303588\\
0.9979	2303444\\
0.9981	2303428\\
0.999	2303428\\
};
\addlegendentry{KGraph}

\addplot+[ultra thick]
  table[row sep=crcr]{%
0.4133	2002192\\
0.4369	4391900\\
0.4631	7412648\\
0.5309	1992392\\
0.5611	4392028\\
0.5843	7412652\\
0.649	1992296\\
0.6835	4391964\\
0.7076	7412652\\
0.7904	1992232\\
0.8171	4391900\\
0.8347	7412780\\
0.872	1992232\\
0.8883	4391836\\
0.9009	7412780\\
0.9269	1992296\\
0.9375	4391836\\
0.9436	7412716\\
0.9671	1992232\\
0.9751	4391964\\
0.9767	7412716\\
0.9852	1992232\\
0.989	4391900\\
0.9897	7412716\\
0.9926	1992232\\
0.9951	4391900\\
0.9956	7412780\\
0.9974	1992164\\
0.9983	4391900\\
0.9987	7412716\\
0.9989	1992168\\
0.9991	7412716\\
0.9992	4392028\\
0.9994	7412716\\
0.9995	4391832\\
0.9996	1992168\\
};
\addlegendentry{Annoy}

\addplot+[ only marks, mark size=3.0pt, mark=square*] coordinates {( 0.9, 133880)};
\addlegendentry{MP-CP}
\addplot[violet,only marks, mark size=4.0pt, mark=diamond*] coordinates {( 1, 94000)};
\addlegendentry{AMIH}
\addplot[only marks, mark size=4.0pt, mark=oplus*, fill=yellow!80!black,draw=black] coordinates {( 0, 14400)};
\addlegendentry{Dataset}
\end{axis}
\end{tikzpicture}
\end{subfigure}
\caption{Average query time and the memory overhead with respect to the recall rate for single-probe crosspolytope (SP-CP) LSH ($k=16$) , multiprobe crosspolytope (MP-CP) LSH, Annoy, KGraph and AMIH. The memory overhead plots also show the size of dataset (the recall rate of zero for dataset size does not have a meaning). For MP-CP, the optimal value of $k$ is 20 for both 64-bit and 128-bit codes.}
\label{fig::AMIH_vs_ANN_binary}
\end{figure*}
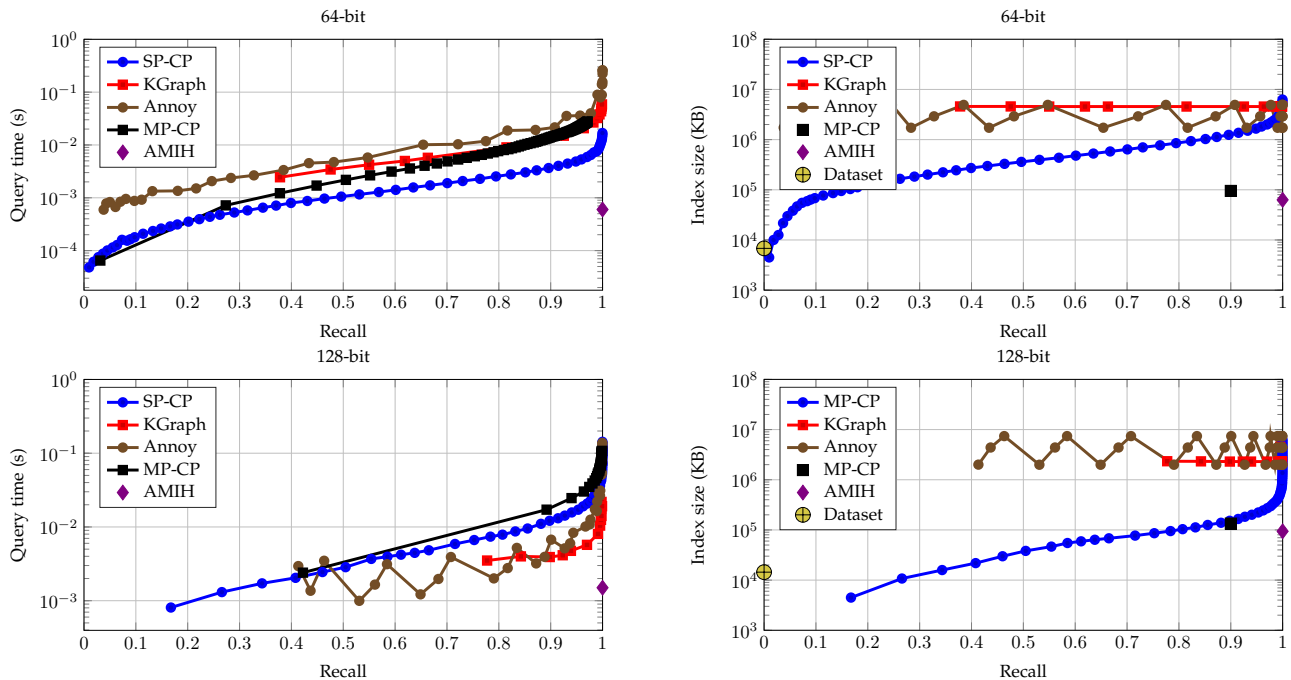

Fig.~\ref{fig::AMIH_vs_ANN_binary} shows the average query time as well as the index size (memory overhead) of each technique with respect to the recall rates. Note that AMIH is an exact algorithm in the binary space therefore its recall rate is 1. The results highlight that AMIH is significantly faster than other techniques for near-perfect recall rates. However, for longer codes the difference between AMIH and other techniques reduces. SP-CP has very fast query time for low recall rates especially in 64-bit codes. In particular, SP-CP is the fastest technique for recall rates smaller than 0.3 in 64-bit codes. The results show that LSH based techniques tend to be faster than  KGraph and Annoy for both lengths of codes. The only exception is in recall rates very close to 1 for which KGraph performs better than other approximate techniques but still slower than AMIH. Another advantage of AMIH over the other techniques is the memory cost. AMIH achieves perfect recall with memory cost that is comparable with the dataset. However, Annoy and KGraph index size can take a large amount of memory, even 100 times more than the size of dataset. Therefore, AMIH is particularly interesting when the RAM budget is quite restrictive. In fact, the high memory cost of Annoy and KGraph did not allow us to provide similar comparisons for the ANN$\_$SIFT1B dataset. The memory cost of Annoy and KGraph remains virtually the same for different recall rates but the catch is that they require more preprocessing time to achieve higher recall rates (the preprocessing time of each technique is not shown here due to limited space).

We would like to note that most of the techniques in ann-benchmark are designed to work with real vectors and may not be necessarily optimized for binary data. Therefore, each of these techniques can be potentially implemented more efficiently to achieve better query time for binary data. However, editing the source code of all techniques in ann-benchmark would require a great deal of human effort and is beyond the scope of this paper.

\subsubsection{Nearest Neighbor Search in Real Space}
If the target binary dataset is generated with binary hashing techniques, then the nearest neighbors found by AMIH are approximate with respect to the original space. For instance, the nearest neighbor found in the above experiments is not exact for the original 128D SIFT vectors. One clear advantage of the approximate techniques such as LSH and KGraph over AMIH is that they are far more general techniques that can work with many distance measures, whereas AMIH is specifically designed for binary spaces. The benefit of applying such approximate techniques in the original space is that they could potentially achieve higher recall rates (with respect to the original space). On the other hand, mapping items to binary codes significantly reduces the storage costs as well as the cost of comparing the items. Therefore, the question that arises is: what is the performance of state-of-the-art approximation techniques in the original space in comparison with AMIH applied to a binary dataset generated by a binary hashing technique?

To answer this question, we compare the performance of SP-CP LSH, MP-CP LSH, KGraph and Annoy applied to the original SIFT vectors with the performance of AMIH applied to the binary vectors generated by AQBC. It is clear that, in this setting, the precision of AMIH is highly dependent on how accurate the binary hash function can preserve the similarities. Learning hash functions to increase accuracy is an active line of research, but is not the focus of this study. Still, such a comparison can be helpful in judging the usefulness of AMIH for non-binary datasets.

Similar to the our previous experiments, we use the ann-benchmark parameter setting for SP-CP, Annoy and KGraph. For MP-CP, we again fix the number of hash tables ($l=10$) and chose the value of $k$ that corresponds to the minimum query time for recall rates above 0.9. For AMIH, we increase $K$ (the number of nearest neighbor to retrieve) from 1 to 1000 and for each value, the $K$NN problem is solved for each query in the binary space. Then, the algorithm linearly scans among the retrieved candidates to find the closest point to the query in the original space. Therefore, the AMIH query time reported in this setting is the summation of: i) the time required to hash the real query point into the binary space (using AQBC), ii) the time to solve $K$NN problem in the binary space with AMIH and iii) the time to perform linear scan among the retrieved points in the original space. This evaluation process of AMIH is similar to the MP-CP. In both, after populating the hash tables, to boost the recall rate, the search algorithm increases the number of probings per hash table in order to retrieve a larger number of candidates. Increasing probings causes better recall rates but also reduces the search speed because we have to probe more buckets and also compare more candidates with the query.

\begin{figure*}
\begin{subfigure}{0.4\linewidth}
\begin{tikzpicture}[scale=0.7]
\begin{axis}[/pgfplots/tick scale binop=\times,
legend pos=north west,
ylabel={Query time (s)},
xlabel={Recall},
ymode=log,
xmajorgrids,
ymajorgrids,
width=4.5in,
height=2.5in,
ymin=0,
xmin=0,xmax=1]

\addplot+[thick] table [x=recall,y=qt, col sep=comma] {amih64.csv};
\addlegendentry{AMIH 64-bit}

\addplot+[thick] table [x=recall,y=qt, col sep=comma] {amih128.csv};
\addlegendentry{AMIH 128-bit}
	
\addplot+[thick] table [x=recall,y=qt, col sep=comma] {kgraph_sift_real.csv};
\addlegendentry{KGraph}

\addplot+[thick] table [x=recall,y=qt, col sep=comma] {annoy_sift_real.csv};
\addlegendentry{Annoy}

\addplot+[thick, green] table [x=recall,y=qt, col sep=comma] {sp-cp_sift_real.csv};
\addlegendentry{SP-CP}

\addplot+[color=violet, thick] table [x=recall,y=qt, col sep=comma] {mp-cp_sift_real.csv};
\addlegendentry{MP-CP}

\end{axis}
\end{tikzpicture}
\end{subfigure}
\quad\quad\quad\quad\quad
\begin{subfigure}{0.4\linewidth}
\begin{tikzpicture}[scale=0.7]
\begin{axis}[/pgfplots/tick scale binop=\times,
legend pos=north west,
ylabel={Index size (KB)},
xlabel={Recall},
ymode=log,
xmajorgrids,
ymajorgrids,
width=4.5in,
height=2.5in,
ymin=0,
xmin=0,xmax=1]
\addplot+[thick]
  table[row sep=crcr]{%
0	56800\\
1	56800\\
};
\addlegendentry{AMIH 64-bit}

\addplot+[thick]
  table[row sep=crcr]{%
0	108400\\
1	108400\\
};
\addlegendentry{AMIH 128-bit}

\addplot+[thick] table [x=recall,y=indexsize, col sep=comma] {kgraph_sift_real.csv};
\addlegendentry{KGraph}

\addplot+[thick] table [x=recall,y=indexsize, col sep=comma] {annoy_sift_real.csv};
\addlegendentry{Annoy}

\addplot+[thick, green] table [x=recall,y=indexsize, col sep=comma] {sp-cp_sift_real.csv};
\addlegendentry{SP-CP}

\addplot+[thick,color=black]
  table[row sep=crcr]{%
0	101124\\
1	101124\\
};
\addlegendentry{MP-CP}

\end{axis}
\end{tikzpicture}
\end{subfigure}
\caption{Average query time and the memory overhead with respect to the recall rate for single-probe crosspolytope (SP-CP) LSH , multiprobe crosspolytope (MP-CP) LSH for $k=20$, Annoy, KGraph and AMIH.}
\label{fig::amih_vs_approx_real}
\end{figure*}
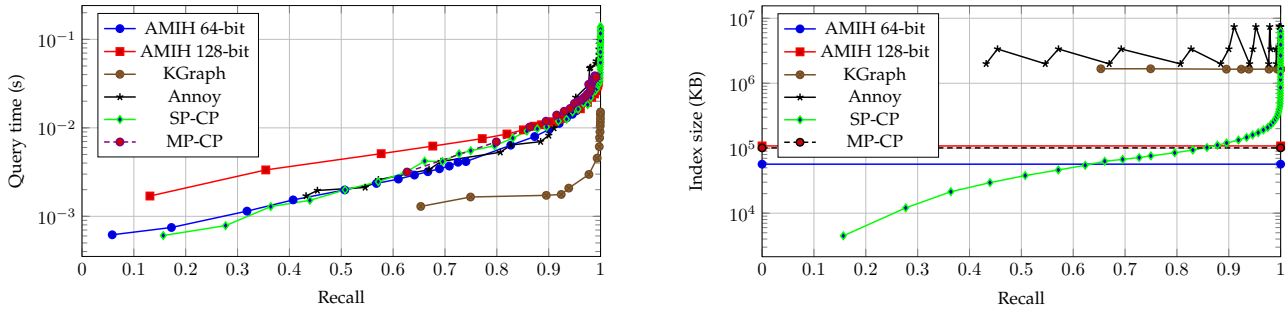
Fig.~\ref{fig::amih_vs_approx_real} shows the average query time and the memory overhead of each technique for the task of angular nearest neighbor search in the real space. In this case, the KGraph clearly outperforms other techniques for all tested recall rates. Among the others, AMIH 64-bit and Annoy show better average query time for many of the recall rate values. For recall rates very close to one, after KGraph, AMIH-128 consistently exhibits the fastest query time. In terms of the memory requirement, AMIH and MP-SP require constant memory budget in all experiments as the number of hash tables remains fixed. The memory footprint of SP-CP increases with recall rate due to the higher number of hash tables. Similar to previous experiments, at low recall rates, SP-CP imposes small memory cost but for recall rates greater than 0.62 AMIH 64-bit has the smallest memory overhead while achieving slightly better recall rates than SP-CP.

We would like to note that the applications of binary hashing or any other approach for compact representation is slightly different from other approximate nearest neighbor search techniques such as KGraph and LSH. Binary hashing techniques are in essence designed for extremely large datasets, too large that we are not even able to store the entire raw dataset in the memory, let alone algorithms that require superlinear storage with large exponents and constants. The goal of binary hashing is to reduce the storage cost of such large datasets in order to fit them in the memory of a single machine while still being faithful to the original metric. Unlike the setting used in ann-benchmark and the experiments of this subsection, in binary hashing applications, not only high memory overheads are not tolerated, but also the datasets itself is often absent in the memory. Moreover, the performance of AMIH with respect to the real space can be improved if more accurate hash functions (than AQBC) are applied to the dataset. Nevertheless, the empirical results of this section show that some approximation techniques with non-compact index structures such as KGraph, perform better when significant memory overhead is not a matter of concern.

\section{Conclusion and Future Work}
This paper proposes a new algorithm for solving the angular $K$NN problem on large-scale datasets of binary codes. By treating binary codes as memory addresses, our proposed algorithm can find similar binary codes in terms of cosine similarity in a time that grows sublinearly with dataset size. To achieve this, we first show a connection between the Hamming distance and the cosine similarity. This connection is in turn used to solve the angular $K$NN problem for applications where binary codes are used as the memory addresses of a hash table. 

Most of our effort were focused on finding nearest neighbors of a binary code in a hash table when codes are compared in terms of the cosine similarity. However, other measures of similarity have been also proposed for different applications. One potential avenue for future work is to find fast search algorithms for other measures of similarity such as the spherical Hamming distance~\cite{heo2012spherical} and the weighted Hamming distance~\cite{zhang2013binary}.  

\balance

\bibliographystyle{plain}
\bibliography{ref}

\begin{thebibliography}{10}

\bibitem{alahi2012freak}
Alexandre Alahi, Raphael Ortiz, and Pierre Vandergheynst.
\newblock Freak: Fast retina keypoint.
\newblock {\em IEEE Conference on Computer Vision and Pattern Recognition},
  pages 510--517, 2012.

\bibitem{andoni2015practical}
Alexandr Andoni, Piotr Indyk, Thijs Laarhoven, Ilya Razenshteyn, and Ludwig
  Schmidt.
\newblock Practical and optimal lsh for angular distance.
\newblock {\em Advances in Neural Information Processing Systems}, pages
  1225--1233, 2015.

\bibitem{aumuller2017ann}
Martin Aum{\"u}ller, Erik Bernhardsson, and Alexander Faithfull.
\newblock Ann-benchmarks: A benchmarking tool for approximate nearest neighbor
  algorithms.
\newblock {\em International Conference on Similarity Search and Applications},
  pages 34--49, 2017.

\bibitem{babenko2012inverted}
Artem Babenko and Victor Lempitsky.
\newblock The inverted multi-index.
\newblock {\em IEEE Conference on Computer Vision and Pattern Recognition},
  pages 3069--3076, 2012.

\bibitem{bayardo2007scaling}
Roberto~J Bayardo, Yiming Ma, and Ramakrishnan Srikant.
\newblock Scaling up all pairs similarity search.
\newblock In {\em International Conference on World Wide Web}, pages 131--140.
  ACM, 2007.

\bibitem{Bentley:1975:MBS:361002.361007}
Jon~Louis Bentley.
\newblock Multidimensional binary search trees used for associative searching.
\newblock {\em Communications of the ACM}, 18(9):509--517, 1975.

\bibitem{annoy}
Erik Bernhardsson.
\newblock Annoy github.com/spotify/annoy.

\bibitem{calonder2010brief}
Michael Calonder, Vincent Lepetit, Christoph Strecha, and Pascal Fua.
\newblock Brief: Binary robust independent elementary features.
\newblock {\em European conference on computer vision}, pages 778--792, 2010.

\bibitem{carr2015}
Miguel~A. Carreira-Perpinan and Ramin Raziperchikolaei.
\newblock Hashing with binary autoencoders.
\newblock {\em IEEE Conference on Computer Vision and Pattern Recognition},
  2015.

\bibitem{dean2013fast}
Thomas Dean, Mark~A Ruzon, Mark Segal, Jonathon Shlens, Sudheendra
  Vijayanarasimhan, and Jay Yagnik.
\newblock Fast, accurate detection of 100,000 object classes on a single
  machine.
\newblock {\em IEEE Conference on Computer Vision and Pattern Recognition},
  pages 1814--1821, 2013.

\bibitem{do2016learning}
Thanh-Toan Do, Anh-Dzung Doan, and Ngai-Man Cheung.
\newblock Learning to hash with binary deep neural network.
\newblock {\em European Conference on Computer Vision}, pages 219--234, 2016.

\bibitem{kgraph}
Wei Dong.
\newblock Kgraph github.com/aaalgo/kgraph.

\bibitem{EghbaliAT17}
Sepehr Eghbali, Hassan Ashtiani, and Ladan Tahvildari.
\newblock Online nearest neighbor search in binary space.
\newblock In {\em {IEEE} International Conference on Data Mining}, pages
  853--858, 2017.

\bibitem{flum2006parameterized}
J.~Flum and M.~Grohe.
\newblock {\em Parameterized Complexity Theory}.
\newblock Springer-Verlag Berlin Heidelberg, 2006.

\bibitem{Friedman:1977:AFB:355744.355745}
Jerome~H. Friedman, Jon~Louis Bentley, and Raphael~Ari Finkel.
\newblock An algorithm for finding best matches in logarithmic expected time.
\newblock {\em ACM Transactions on Mathmatical Software}, 3(3):209--226, 1977.

\bibitem{Gong_angular:NIPS2012_4831}
Yunchao Gong, Sanjiv Kumar, Vishal Verma, and Svetlana Lazebnik.
\newblock Angular quantization-based binary codes for fast similarity search.
\newblock {\em Advances in Neural Information Processing Systems}, pages
  1196--1204, 2012.

\bibitem{gong2013iterative}
Yunchao Gong, Svetlana Lazebnik, Albert Gordo, and Florent Perronnin.
\newblock Iterative quantization: A procrustean approach to learning binary
  codes for large-scale image retrieval.
\newblock {\em IEEE Transactions on Pattern Analysis and Machine Intelligence},
  35(12):2916--2929, 2013.

\bibitem{greene1994multi}
Dan Greene, Michal Parnas, and Frances Yao.
\newblock Multi-index hashing for information retrieval.
\newblock {\em Annual Symposium on Foundations of Computer Science}, pages
  722--731, 1994.

\bibitem{hajishirzi2010adaptive}
Hannaneh Hajishirzi, Wen-tau Yih, and Aleksander Kolcz.
\newblock Adaptive near-duplicate detection via similarity learning.
\newblock {\em Proceedings of the 33rd international ACM SIGIR conference on
  Research and development in information retrieval}, pages 419--426, 2010.

\bibitem{heo2012spherical}
Jae-Pil Heo, Youngwoon Lee, Junfeng He, Shih-Fu Chang, and Sung-Eui Yoon.
\newblock Spherical hashing.
\newblock {\em International Conference on Computer Vision and Pattern
  Recognition}, pages 2957--2964, 2012.

\bibitem{Indyk:1998:ANN:276698.276876}
Piotr Indyk and Rajeev Motwani.
\newblock Approximate nearest neighbors: Towards removing the curse of
  dimensionality.
\newblock {\em Annual ACM Symposium on Theory of Computing}, pages 604--613,
  1998.

\bibitem{iwamura2013most}
Mikio Iwamura, Takao Sato, and Kenji Kise.
\newblock What is the most efficientway to select nearest neighbor candidates
  for fast approximate nearest neighbor search?
\newblock {\em IEEE International Conference on Computer Vision}, pages
  3535--3542, 2013.

\bibitem{product}
H.~Jegou, M.~Douze, and C.~Schmid.
\newblock Product quantization for nearest neighbor search.
\newblock {\em IEEE Transactions on Pattern Analysis and Machine Intelligence},
  33(1):117--128, 2011.

\bibitem{kong2012manhattan}
Weihao Kong, Wu-Jun Li, and Minyi Guo.
\newblock Manhattan hashing for large-scale image retrieval.
\newblock {\em International Conference on Research and Development in
  Information Retrieval}, pages 45--54, 2012.

\bibitem{kulis2009learning}
Brian Kulis and Trevor Darrell.
\newblock Learning to hash with binary reconstructive embeddings.
\newblock {\em Advances in Neural Information Processing Systems}, pages
  1042--1050, 2009.

\bibitem{leutenegger2011brisk}
Stefan Leutenegger, Margarita Chli, and Roland~Y Siegwart.
\newblock Brisk: Binary robust invariant scalable keypoints.
\newblock {\em International Conference on Computer Vision}, pages 2548--2555,
  2011.

\bibitem{Liu_2016_CVPR}
Haomiao Liu, Ruiping Wang, Shiguang Shan, and Xilin Chen.
\newblock Deep supervised hashing for fast image retrieval.
\newblock {\em IEEE Conference on Computer Vision and Pattern Recognition},
  2016.

\bibitem{liu2012supervised}
Wei Liu, Jun Wang, Rongrong Ji, Yu-Gang Jiang, and Shih-Fu Chang.
\newblock Supervised hashing with kernels.
\newblock {\em IEEE Conference on Computer Vision and Pattern Recognition},
  pages 2074--2081, 2012.

\bibitem{liu2013reciprocal}
Xianglong Liu, Junfeng He, and Bo~Lang.
\newblock Reciprocal hash tables for nearest neighbor search.
\newblock {\em American Association for Artificial Intelligence}, 2013.

\bibitem{lv2007multi}
Qin Lv, William Josephson, Zhe Wang, Moses Charikar, and Kai Li.
\newblock Multi-probe lsh: efficient indexing for high-dimensional similarity
  search.
\newblock {\em International Conference on Very Large Data Bases}, pages
  950--961, 2007.

\bibitem{mao2016two}
Minqi Mao, Zhonglong Zheng, Zhongyu Chen, Huawen Liu, Xiaowei He, and Ronghua
  Ye.
\newblock Two-dimensional pca hashing and its extension.
\newblock {\em International Conference on Pattern Recognition}, pages
  1624--1629, 2016.

\bibitem{aizawa1pqtable}
Yusuke Matusi, Toshihiko Yamasaki, and Kiyoharu Aziawa.
\newblock Pqtable: Fast exact asymmetric distance neighbor search for product
  quantization using hash tables.
\newblock {\em International Conference on Computer Vision}, 2015.

\bibitem{muja2012fast}
Marius Muja and David~G Lowe.
\newblock Fast matching of binary features.
\newblock {\em Conference on Computer and Robot Vision}, pages 404--410, 2012.

\bibitem{norouzi2011minimal}
Mohammad Norouzi and David~M Blei.
\newblock Minimal loss hashing for compact binary codes.
\newblock {\em International Conference on Machine Learning}, pages 353--360,
  2011.

\bibitem{norouzi2012fast}
Mohammad Norouzi, Ali Punjani, and David~J Fleet.
\newblock Fast search in hamming space with multi-index hashing.
\newblock {\em IEEE Conference on Computer Vision and Pattern Recognition},
  pages 3108--3115, 2012.

\bibitem{norouzi2014fast}
Mohammad Norouzi, Ali Punjani, and David~J Fleet.
\newblock Fast exact search in hamming space with multi-index hashing.
\newblock {\em IEEE Transactions on Pattern Analysis and Machine Intelligence},
  36(6):1107--1119, 2014.

\bibitem{ong2016improved}
Eng-Jon Ong and Miroslaw Bober.
\newblock Improved hamming distance search using variable length substrings.
\newblock {\em IEEE Conference on Computer Vision and Pattern Recognition},
  pages 2000--2008, 2016.

\bibitem{Ong_2016_CVPR}
Eng-Jon Ong and Miroslaw Bober.
\newblock Improved hamming distance search using variable length substrings.
\newblock {\em IEEE Conference on Computer Vision and Pattern Recognition},
  2016.

\bibitem{localitycomparison}
Loïc Paulevé, Hervé Jégou, and Laurent Amsaleg.
\newblock Locality sensitive hashing: A comparison of hash function types and
  querying mechanisms.
\newblock {\em Pattern Recognition Letters}, 31(11):1348 -- 1358, 2010.

\bibitem{falconn}
Ilya Razenshteyn and Ludwig Schmidt.
\newblock Fast lookups of cosine and other nearest neighbors
  github.com/falconn-lib/falconn.

\bibitem{rublee2011orb}
Ethan Rublee, Vincent Rabaud, Kurt Konolige, and Gary Bradski.
\newblock Orb: An efficient alternative to sift or surf.
\newblock {\em International conference on computer vision}, pages 2564--2571,
  2011.

\bibitem{salakhutdinov2009semantic}
Ruslan Salakhutdinov and Geoffrey Hinton.
\newblock Semantic hashing.
\newblock {\em International Journal of Approximate Reasoning}, 50(7):969--978,
  2009.

\bibitem{samet2006foundations}
Hanan Samet.
\newblock {\em Foundations of multidimensional and metric data structures}.
\newblock Morgan Kaufmann, 2006.

\bibitem{shakhnarovich2003fast}
Gregory Shakhnarovich, Paul~A Viola, and Trevor Darrell.
\newblock Fast pose estimation with parameter-sensitive hashing.
\newblock {\em ICCV}, 3:750, 2003.

\bibitem{shrivastava2014defense}
Anshumali Shrivastava and Ping Li.
\newblock In defense of minhash over simhash.
\newblock {\em International Conference on Artificial Intelligence and
  Statistics}, pages 886--894, 2014.

\bibitem{wang2014hashing}
Jingdong Wang, Heng~Tao Shen, Jingkuan Song, and Jianqiu Ji.
\newblock Hashing for similarity search: A survey.
\newblock {\em arXiv preprint arXiv:1408.2927}, 2014.

\bibitem{weber1998quantitative}
Roger Weber, Hans-J{\"o}rg Schek, and Stephen Blott.
\newblock A quantitative analysis and performance study for similarity-search
  methods in high-dimensional spaces.
\newblock {\em International Conference on Very Large Data bases}, 98:194--205,
  1998.

\bibitem{weiss2009spectral}
Yair Weiss, Antonio Torralba, and Rob Fergus.
\newblock Spectral hashing.
\newblock {\em Advances in Neural Information Processing Systems}, pages
  1753--1760, 2009.

\bibitem{zhang2013binary}
Lei Zhang, Yongdong Zhang, Jinhu Tang, Ke~Lu, and Qi~Tian.
\newblock Binary code ranking with weighted hamming distance.
\newblock {\em IEEE Conference on Computer Vision and Pattern Recognition},
  pages 1586--1593, 2013.

\bibitem{zhang2013topology}
Lei Zhang, Yongdong Zhang, Jinhui Tang, Xiaoguang Gu, Jintao Li, and Qi~Tian.
\newblock Topology preserving hashing for similarity search.
\newblock {\em International Conference on Multimedia}, pages 123--132, 2013.

\bibitem{Zhang_2016_CVPR}
Ziming Zhang, Yuting Chen, and Venkatesh Saligrama.
\newblock Efficient training of very deep neural networks for supervised
  hashing.
\newblock {\em IEEE Conference on Computer Vision and Pattern Recognition},
  2016.

\end{thebibliography}
\end{document}